\documentclass[aps,prd,twocolumn,showpacs,superscriptaddress,groupedaddress,secnum,subsecnum]{revtex4}
\usepackage{graphicx}  
\usepackage{dcolumn}   
\usepackage{bm}        
\usepackage{amssymb}   
\usepackage[unicode=true,colorlinks=true,linkcolor=magenta, urlcolor=blue, citecolor = blue,breaklinks]{hyperref}
\usepackage{multirow}
\usepackage{url}
\usepackage{breakurl}
\usepackage{amsmath}
\usepackage{placeins} 
\usepackage{appendix}

\begin{document}

\def\addressa{Faculty of Physics, Lomonosov Moscow State University, Russia, Moscow, 119234}
\def\addressb{Skobeltsyn Institute of Nuclear Physics, Lomonosov Moscow State University, Russia, Moscow, 119234}
\def\addressc{Moscow Institute of Physics and Technology, Russia, Moscow, 141701}

\title{Coulomb corrections in rare decays of neutral $B$ mesons with $\ell^+\ell^-$-pair in final state}
\author{\firstname{S.I.}~\surname{Manukhov}\footnote{Email: manuhov2000akk@gmail.com }}
\affiliation{\addressa}
\affiliation{\addressb}
\author{\firstname{N.V.}~\surname{Nikitin}}
\affiliation{\addressa}
\affiliation{\addressb}
\affiliation{\addressc}

\date{\today}

\begin{abstract}
We present a systematic analysis of Coulomb corrections for leptonic ($B^0_{d,s}\to \ell^+\ell^-$), semileptonic ($B^0_{d,s}\to h^0\,\ell^+\ell^-$, $B^0_{d,s}\to V^0\ell^+\ell^-$) and radiative leptonic ($B^0_{d,s}\to \gamma \ell^+\ell^-$) decays of neutral $B$-mesons. The relativization of the Coulomb factor was performed by comparing the Gamow-Sommerfeld-Sakharov factor, the exact relativistic approach of Crater-Alstine-Sazdjian applied by us to scalar systems, and well-known one-loop QED calculations. Coulomb corrections are calculated for differential, angular, and double-differential distributions, as well as for partial decay widths. We also discuss the role of the Coulomb factor among other QED corrections, in particular, the contribution of soft-photon radiation, which is effectively simulated in experiments by tools such as \textsc{photos}.

For the $B_s^0 \to \mu^+\mu^-$ channel, Coulomb corrections improve the prediction of the partial width to $\delta = |\mathcal{B}^{(exp)} - \mathcal{B}^{(theory)}|/\mathcal{B}^{(exp)} = 2\%$. This improvement brings the prediction closer to the LHCb/CMS experimental results within the current experimental (11\%) and theoretical (5\% lattice QCD) errors.  In the decays $B^0\to K^0\mu^+\mu^-$ and $B^0 \to K^{0*}\mu^+\mu^-$, Coulomb effects also reduce the discrepancies between theoretical predictions and experimental data (from $\delta = 2\%$ to less than $\delta = 1\%$ and from $\delta = 11\%$ to $\delta = 4\%$ respectively). Finally, for the decays involving $\tau$-leptons, the Coulomb correction reaches $4\%$. While currently smaller than the dominant form-factor uncertainties and experimental errors, the Coulomb correction represents a non-negligible systematic effect. It should be accounted for in the high-precision era of $B$-physics, where such effects may become significant for the interpretation of potential New Physics signals. 
\end{abstract}
\par

\maketitle
\section{Introduction}\label{intro}

Rare semileptonic and ultra-rare leptonic decays of $B$ mesons are under intensive investigation at the LHCb \cite{LHCb:2024onj,LHCb:2024uff,LHCb:2022qnv,LHCb:2021trn,PhysRevLett.128.041801,LHCb:2021awg,LHCb:2017rmj,LHCb:2016ykl,LHCb:2014cxe, LHCb:2021zwz}, CMS \cite{CMS:2024atz,CMS:2024syx,CMS:2022mgd}, and ATLAS \cite{ATLAS:2018cur} experiments at the Large Hadron Collider, as well as in the Belle-II experiment \cite{Belle:2024cis,PhysRevLett.131.051804,Belle-II:2022fky}. At present, the partial widths of the leptonic decays $B^0_{d,s}\to \mu^+\mu^-$ have been measured \cite{LHCb:2021awg,ATLAS:2018cur,CMS:2022mgd}, along with the semileptonic decays $B_{d,s}^0\to X^0 \,\ell^+\ell^-$ (where $X^0 = {K^{0(*)}, \eta,\phi,\omega,\pi^0,\rho^0, \ldots}$ is a pseudoscalar or vector meson) \cite{LHCb:2024onj,Belle:2024cis}. Upper limits have also been established for the leptonic radiative decay $B^0_s\to \gamma \ell^+\ell^-$ \cite{LHCb:2024uff}. Differential and angular distributions have been obtained for the decays $B^0\to K^{0(*)}\mu^+\mu^-$ and $B_s^0\to \phi\,\mu^+\mu^-$ \cite{CMS:2024atz,LHCb:2024onj,LHCb:2016ykl,LHCb:2014cxe, LHCb:2021zwz}.

These decays have been theoretically studied in detail both within the Standard Model (SM) and in its extensions. Some works on this broad topic can be found in Refs. \cite{Buras:1995iy, Buchalla:1995vs, Buras:2001pn, Beneke:2019slt,PhysRevD.57.6814,PhysRevD.70.114028,PhysRevD.101.096007, PhysRevLett.120.011801, Huang:2023nli, PhysRevD.108.L031502, Isidori:2020acz, Buras:2012ru, Cali:2019nwp, Bigi:2023cbv, Isidori:2022bzw}. One of the most precisely measured quantities is the partial width of the $B_s^0\to\mu^+\mu^-$ decay, determined by the ATLAS \cite{ATLAS:2018cur}, CMS \cite{CMS:2022mgd}, and LHCb \cite{PhysRevLett.128.041801,LHCb:2017rmj} collaborations. While early measurements (before 2022) tended to lie below the SM prediction~\cite{Beneke:2019slt}, the latest CMS result~\cite{CMS:2022mgd} shows a central value slightly above it, with theory and experiment remaining consistent within the current uncertainties.

A different picture emerges from the differential distributions of $B^0\to K^0 \mu^+\mu^-$ \cite{LHCb:2014cxe}, $B^0\to K^{0*} \mu^+\mu^-$ \cite{LHCb:2016ykl, LHCb:2024onj}, and $B_s^0\to \phi\,\mu^+\mu^-$ \cite{LHCb:2013tgx, LHCb:2021zwz}. For $B^0\to K^0 \mu^+\mu^-$, the LHCb analysis \cite{LHCb:2014cxe} shows that the SM predictions lie systematically above the measured branching fraction at low $q^2 = (p_B-p_{K})^2$. The situation is more complex for $B^0\to K^{*0} \mu^+\mu^-$. The 2016 LHCb analysis \cite{LHCb:2016ykl} also indicated a deficit at low $q^2$, but the recent 2024 work \cite{LHCb:2024onj} finds that the differential observables are compatible with the SM within uncertainties. However, a mild tension remains in the Wilson coefficient $C_9$, which exhibits a $2.1\sigma$ deviation from the SM prediction. Another interesting difference appears in the $B_s^0\to \phi \mu^+\mu^-$ channel. Already the first measurement in 2013 suggested a deficit~\cite{LHCb:2013tgx}. The latest LHCb results (2021) confirm a persistent discrepancy of about $3.6\sigma$~\cite{LHCb:2021zwz}.

The existing discrepancies between theoretical predictions and experimental data  motivate more precise calculations of various QCD and QED corrections.  For the $B^0_{s,d}\to\mu^+\mu^-$ decays, a systematic calculation of the leading  logarithmic QED corrections and the mixed QED--QCD corrections (hard-collinear and soft) was performed in Ref.~\cite{Beneke:2019slt} within the framework of a soft-collinear effective theory (SCET); this work is often used as a benchmark by experimental collaborations (e.g.,~\cite{PhysRevLett.128.041801, CMS:2022mgd}). Physically, the dominant QED correction originates from the interaction of a final-state lepton with the spectator quark inside the $B$ meson. Such interactions give rise to power-enhanced logarithms: the large logarithms $\ln(m_b/\Lambda_{\text{QCD}})$ come multiplied by a factor $m_b/\Lambda_{\text{QCD}}$, making the radiative corrections numerically much larger than expected from the $\alpha_{\text{em}}$ suppression alone. We note that this effect was first identified in~\cite{PhysRevLett.120.011801} and was later extended to other decays, such as $B^0_{s,d}\to\tau^+\tau^-$ \cite{Huang:2023nli} and $B^-\to\mu^-\bar\nu_\mu$~\cite{PhysRevD.108.L031502}.

In contrast to the structure-dependent effects analysed in Ref.~\cite{Beneke:2019slt}, a complementary line of research addresses structure-independent QED corrections. For the semileptonic decays $B\to K\ell^+\ell^-$, such corrections were evaluated in Ref.~\cite{Isidori:2020acz} using the slicing method, where the impact of soft and hard-collinear photon radiation is accounted for at the double-differential level, while the mesons are treated as pointlike particles.

On general grounds, it is well understood that only a suitably defined decay rate  $\Gamma = \Gamma(B^0_{d,s}\to X^0\ell^+\ell^-) + \sum_\gamma 
\Gamma(B^0_{d,s}\to X^0\ell^+\ell^-\gamma)\,\theta(\Delta E-\sum E_\gamma)$,  in which the non-radiative part includes virtual corrections, is infrared-finite and well-defined. The result retains a dependence on the photon-energy cutoff  $\Delta E$ imposed by the experimental setup, which requires the inclusion of  an arbitrary number of undetected real photons with energy $E_\gamma < \Delta E$  in the theoretical prediction. The soft-photon emission from the final-state leptons is currently simulated in experimental analyses with tools like  \textsc{photos}~\cite{Barberio:1993qi,Golonka:2005pn}, such that the measured branching fraction is interpreted as the non-radiative one~\cite{Buras:2012ru}.

While \textsc{photos} accounts for real soft radiation, it does not model virtual photon exchanges, and in particular does not include the Coulomb interaction that arises between two charged final-state particles (see Ref.~\cite{Cali:2019nwp} for a demonstration in $\bar{B}^0 \to D^+ \ell^- \bar{\nu}_\ell$ decays). The Coulomb correction has also been addressed in the recent comprehensive analysis of QED effects in inclusive $B \to X_c \ell\nu$ decays~\cite{Bigi:2023cbv}. That work includes, among many other contributions, the Coulomb interaction between the final-state charm quark and lepton, and explicitly retains it even far from the threshold region.

Given the importance of the Coulomb effect, one may ask whether it has been taken into account in the rare decays of neutral $B$ mesons with an $\ell^+\ell^-$ pair. Despite the comprehensive treatment of QED corrections in Refs.~\cite{Beneke:2019slt} and \cite{Isidori:2020acz}, neither analysis incorporates the Coulomb interaction between the charged final-state leptons. The work of Ref.~\cite{Beneke:2019slt}, as discussed, concentrates on structure-dependent QED--QCD effects; its soft-radiation contribution (Eq.~(8.13)) reduces to the universal factor simulated by \textsc{photos}, so no lepton--lepton Coulomb interaction is present. The calculation of Ref.~\cite{Isidori:2020acz} has been shown to be numerically equivalent to \textsc{photos}~\cite{Isidori:2022bzw}, again omitting Coulomb interaction between the final-state particles. Thus, the Coulomb correction is still missing from the literature on neutral $B$-meson decays. 

In this paper we fill this gap by systematically accounting for the Coulomb interaction between charged leptons in the final state of neutral $B$-meson decays. In Section \ref{sec2}, several approaches for calculating Coulomb corrections are compared: the non-relativistic Gamow-Sommerfeld-Sakharov (GSS) method, the relativistic Crater-Alstine-Sazdjian (CAS) formalism based on exact two-particle relativistic equations, and the approach based on QED loop calculations. The relation between the Coulomb correction and soft-photon radiation is also addressed. In Section \ref{sec3}, the Coulomb correction is applied for the analysis of ultra-rare leptonic decays $B^0_{d,s}\to \ell^+\ell^-$; in Section \ref{sec4}, for the analysis of rare semileptonic decays $B_{d,s}^0\to h^0\ell^+\ell^-$ with a pseudoscalar meson $h^0$; in Section \ref{sec5}, for the analysis of rare semileptonic decays $B_{d,s}^0\to V^0\ell^+\ell^-$ with a vector meson $V^0$; and in Section \ref{sec6}, for the analysis of rare radiative semileptonic decays $B_{d,s}^0\to \gamma\ell^+\ell^-$. 

\section{Role of the Coulomb interaction in QED corrections and methods for its treatment}
\label{sec2}

The Coulomb interaction is described by the $\mathcal{K}_C$-factor:

\begin{equation}
\mathcal{K}_\text{C} = \frac{\mathcal{B}^{(\text{Coulomb})}}{\mathcal{B}^{(\text{free})}},
\end{equation}
where $\mathcal{B}^{(\text{Coulomb})}$ denotes the branching fraction with the Coulomb interaction in the final state, and $\mathcal{B}^{(\text{free})}$ denotes the branching fraction without such interaction.

The $\mathcal{K}_C$-factor can be calculated using several methods. To validate the approach for neutral $B$-meson decays, we first benchmark these methods against a simpler process: the decay of a hypothetical neutral scalar particle $B^0(M)$ into two hypothetical charged scalars $S^+(m)$ and $S^-(m)$. In Subsection~\ref{GSS_sec}, we discuss the approach of Gamow, Sommerfeld, and Sakharov, while in Subsection~\ref{CAS_sec} we apply the relativistic two-particle equation formalism developed by Crater, Alstine, and Sazdjian. In Subsection~\ref{loop_qed}, we place the Coulomb correction in the context of the complete QED soft-photon resummation, discuss its factorisation to soft $\mathcal{K}_{\text{soft}}$ and Coulomb $\mathcal{K}_\text{C}$ terms. 

\subsection{Gamow-Sommerfeld-Sakharov (GSS) Method}
\label{GSS_sec}
\begin{figure}[t]
\begin{center}
\includegraphics[width=92mm]{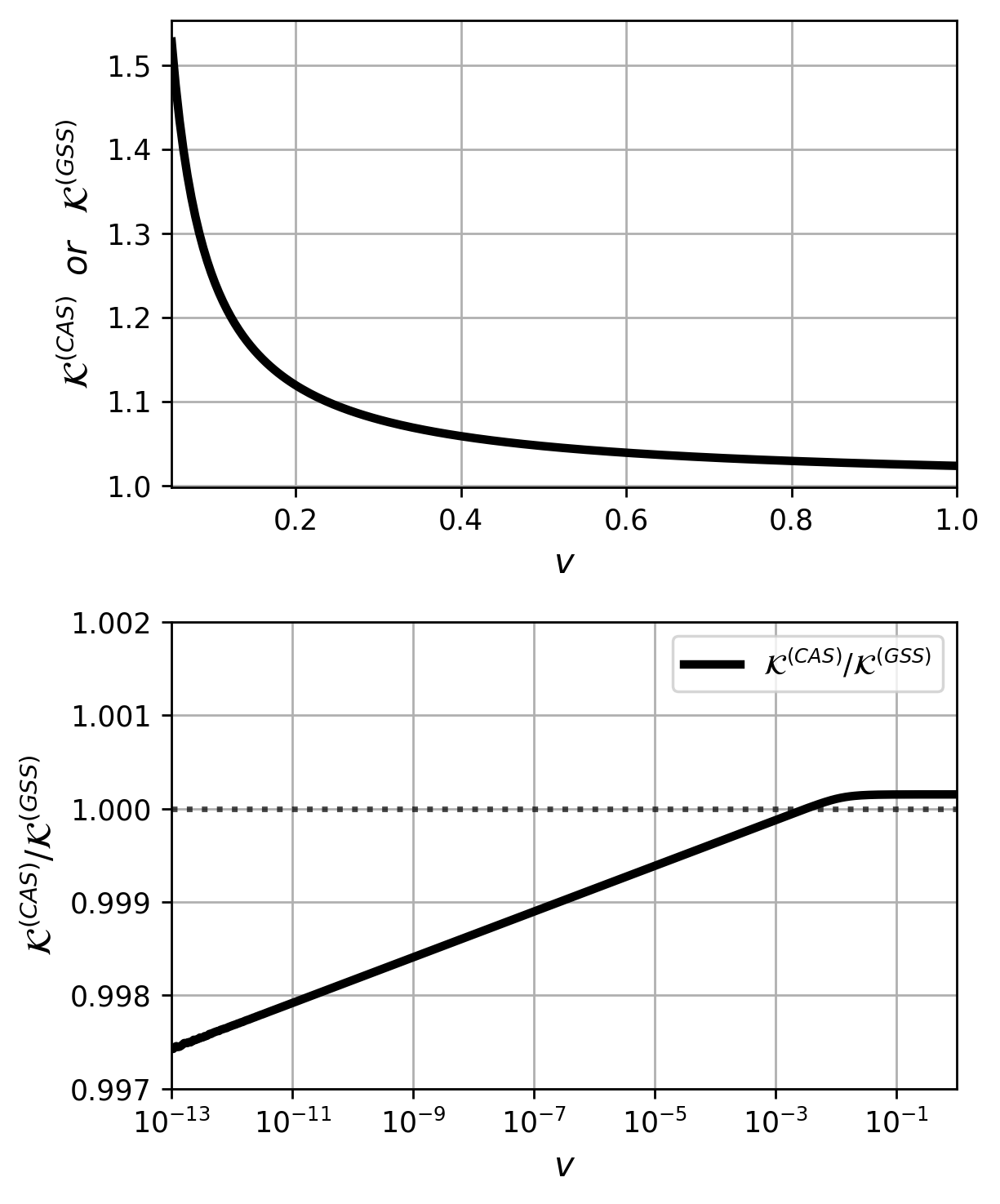}
\vspace{-3mm}
\caption{The dependence of the Coulomb $\mathcal{K}_\text{C}$-factor as a function of the relative velocity $v = v_{rel}$ (top), and the ratio $\mathcal{K}^{(CAS)}/\mathcal{K}^{(GSS)}$ of the factors as a function of $v$ (bottom). The smaller the velocity, the larger the Coulomb enhancement. It can be seen that the CAS and GSS methods yield practically identical results over the entire range of significant velocities $v\in[10^{-13}, 1]$. The relative difference across considered range of velocities does not exceed $0.3\%$.}
\label{pic_Gamov_Crater_Furry_comparison}
\end{center}
\vspace{-5mm}
\end{figure}
The Gamow–Sommerfeld–Sakharov coefficient \cite{Gamow1928,sommerfeld1921atombau,Sakharov:1991} accounts for the Coulomb interaction at \textit{non-relativistic} velocities. This correction, obtained from the Schrödinger equation for the relative motion, takes the form:

\begin{equation}
\label{GSS_factor}
\mathcal{K}_\text{C}^{(\text{GSS})}(v) = \frac{2\pi\alpha_{\text{em}}/v}{1-e^{-2\pi \alpha_{\text{em}}/v}},
\end{equation}
where $v = v_{rel}$ is the relative velocity of particles $S^+$ and $S^-$, and $\alpha_{\text{em}}\approx 1/137$ is the fine structure constant.

\subsection{Crater-Alstine-Sazdjian (CAS) Formalism}
\label{CAS_sec}

The Crater-Alstine-Sazdjian formalism \cite{Crater:1983ew, Crater:1994rt, Sazdjian:1985pg} is a method based on constructing exact relativistic two-particle equations. This approach utilizes classical relativistic dynamics with constraints followed by quantization. A brief overview of the CAS approach using the example of scalar particles is provided in Appendix \ref{app1}. Applying this formalism to the scalar decay leads to the following Coulomb correction (proof is given in Appendix \ref{CAS_derivation}):

\begin{equation}
\label{eq:CAS_factor}
\mathcal{K}_\text{C}^{(\text{CAS})}(v) = \left| \frac{\Gamma\left(\sqrt{\frac{1}{4} - \alpha_{\text{em}}^2} + \frac{1}{2} + i\frac{\alpha_{\text{em}}}{v}\right)}{\Gamma\left(\sqrt{1 - 4\alpha_{\text{em}}^2} + 1\right)} \right|^2 \cdot e^{\pi\alpha_{\text{em}}/v},
\end{equation}
where $\Gamma(x)$ is Euler's gamma function.

The dependence of the $\mathcal{K}_\text{C}$-factor on the relative velocity $v$ of the $S^+S^-$ pair is shown in Fig.~\ref{pic_Gamov_Crater_Furry_comparison}, top. It can be seen that the smaller the final relative velocity, the larger the magnitude of the Coulomb correction. The CAS and GSS corrections yield practically identical results over the entire physically significant velocity range. The ratio of these factors is shown in Fig.~\ref{pic_Gamov_Crater_Furry_comparison}, bottom.

The numerical difference between the CAS and GSS corrections does not exceed 0.3\% in the non-relativistic limit ($v = 10^{-15}c$):
\begin{equation}
\frac{\mathcal{K}_\text{C}^{(\mathrm{CAS})}}{\mathcal{K}_\text{C}^{(\mathrm{GSS})}}\Bigg|_{v = 10^{-15}c} \approx 0.997.
\end{equation}
In the relativistic regime ($v \approx c$), the discrepancy between the methods becomes even smaller:
\begin{equation}
    \lim\limits_{v\to c}\frac{\mathcal{K}_\text{C}^{(CAS)}}{\mathcal{K}_\text{C}^{(GSS)}} \approx 1.0002.
\end{equation}

Thus, the two methods are numerically equivalent with the required precision,  $\mathcal{K}_\text{C}^{(\mathrm{GSS})}\approx\mathcal{K}_\text{C}^{(\mathrm{CAS})}$. Thus, we will denote the Coulomb correction simply by $\mathcal{K}_\text{C}$ and employ the CAS result in the actual calculations.

\subsection{Soft-photon resummation and the Coulomb term}
\label{loop_qed}

In quantum field theory, the Coulomb correction naturally emerges as one of the parts of the soft photon QED correction \cite{Isidori:2007zt}. Adapting the formulas for a decay $B^0(M)\to S^+(m)S^-(m)$, we have:
\begin{widetext}
\begin{equation}
\label{Isidori_factor}
\begin{split}
    \mathcal{K}_{QED} = &\; \mathcal{K}_\text{soft}\cdot \mathcal{K}_\text{C}\cdot \left[ 1- \frac{\alpha}{\pi} (F+ H^{\text{IR}})\right], \text{ where} \\
    \mathcal{K}_\text{soft} = &\; \left(\frac{2\Delta E}{M}\right)^{\tfrac{2\alpha_{\text{em}}}{\pi}\cdot\tfrac{1}{4v}\ln\left(\tfrac{1+v}{1-v}\right)}; \quad\quad  \mathcal{K}_\text{C} =  1 + \frac{\pi\alpha_{\text{em}}}{v} + O\left(\alpha_{\text{em}}^2\right)
\end{split}
\end{equation}
\end{widetext}
where $v = v_{\text{rel}}$ is the relative velocity of the final-state scalars and $\Delta E$ is the experimental photon-energy cutoff. The explicit expressions for $F$ and $H^{\text{IR}}$ are given in Ref.~\cite{Isidori:2007zt}; numerically they are of $\mathcal{O}(\tfrac{\alpha_{\mathrm{em}}}{\pi}(F+ H^{\text{IR}}))\sim \mathcal{O}(\tfrac{\alpha_{\mathrm{em}}}{\pi})\sim 0.2\%$ and can be safely neglected in the present analysis.

The one-loop expression for $\mathcal{K}_\text{C}$ in Eq.~(\ref{Isidori_factor}) coincides with the first-order expansion of the Gamow--Sommerfeld--Sakharov (GSS) factor, confirming that the Coulomb correction is part of the soft-photon QED dressing. The relativistic generalisation of the GSS factor has a long history. Two distinct approaches exist in the literature. The first relies on direct loop QED calculations~\cite{Arbuzov:2011ff,PhysRevD.56.7276,Solovtsova:2009zq}. The second uses so-called relativistic quasipotential equations~\cite{PhysRevD.3.2351,Arbuzov:1993qc,Yoon:2004sr}. Both approaches find that the correction retains the GSS form, with the non-relativistic relative velocity simply replaced by its relativistic analogue $v = v_{\text{rel}} = \sqrt{1-4m^2/s}/(1-2m^2/s)$. Our CAS analysis (Subsection~\ref{CAS_sec}) confirms this conclusion, agreeing with GSS with required accuracy.

Another important feature of the Coulomb correction $\mathcal{K}_\text{C}$ should be emphasised. This factor does not reduce to unity even in the relativistic regime $v\to1$. This can be seen from the explicit structure of the virtual correction, which can be decomposed as $H = H^{\text{UV}} + H^{\text{IR}} + H^{\text{C}}$ \cite{Isidori:2007zt}. The ultraviolet part $H^{\text{UV}}$ becomes finite after renormalisation, while the infrared divergence $H^{\text{IR}}$ cancels against the corresponding divergence from real photon emission. Finally, there is a non‑vanishing contribution $H^{\text{C}} = \pi^2/(2v)$, which is directly related to the Coulomb correction. One readily sees that $H^{\text{C}}$ grows as $v\to0$, so the Coulomb correction is particularly important near threshold. However, it remains non-zero (and enhanced by $\pi^2$) even as $v\to1$, i.e., far from threshold.

Thus, the full correction effectively factorises into a soft $\mathcal{K}_\text{soft}$ and a Coulomb $\mathcal{K}_\text{C}$ factors. As discussed in the Introduction, the $\mathcal{K}_\text{soft}$ is effectively captured by Monte Carlo tools such as \textsc{photos}, while the $\mathcal{K}_\text{C}$ is not included in those simulations. In the following sections we apply $\mathcal{K}_\text{C}$ to the analysis of leptonic, semileptonic and radiative leptonic decays of neutral $B$ mesons.

\section{Coulomb Interaction in \texorpdfstring{$B^0_{d,s}\to \ell^+\ell^-$}{B0d,s → ℓ+ℓ-} Decays}
\label{sec3}

This section we consider the decay of a neutral $B$-meson $(P, M)$ into a lepton pair $\ell^+(p_1,m_\ell)\ell^-(p_2,m_\ell)$ accounting for the Coulomb interaction in the final state. 

Neglecting the masses of the light quarks $q$, the effective Hamiltonian for $b \to q \ell^+\ell^-$ transitions ($q=d,s$) is written as a Wilson operator product expansion \cite{Buchalla:1995vs, PhysRevD.52.186}:
\begin{widetext}
\begin{multline}
\label{Effective_Hamiltonian}
    \mathcal{H}_{eff}^{b\to q \, \ell^+\ell^-}(x) = \frac{G_F}{\sqrt{2}} \frac{\alpha_{\text{em}}}{2\pi} V_{tb} V_{tq}^* \Bigg[ -2im_b \frac{C_{7}(\mu)}{q^2} \cdot \bar{q}(x)\sigma_{\mu\nu} q^\nu (1 + \gamma_5) b(x) \cdot \bar{\ell}(x) \gamma^\mu \ell(x) \\
     + C_{9V}(\mu) \cdot \bar{q}(x) \gamma_\mu (1 - \gamma_5) b(x) \cdot \bar{\ell}(x) \gamma^\mu \ell(x) + C_{10A}(\mu) \cdot \bar{q}(x) \gamma_\mu (1 - \gamma_5) b(x) \cdot \bar{\ell}(x) \gamma^\mu \gamma_5 \ell(x) \Bigg] + h.c.,
\end{multline}
\end{widetext}
where \( G_F \) is the Fermi constant, \( V_{tb} \) and \( V_{tq} \) are elements of the Cabibbo-Kobayashi-Maskawa matrix \cite{ParticleDataGroup:2024cfk}, \( q^\nu \) is the 4-momentum of the lepton pair, and \( q^2 = q^\nu q_\nu \). Here, $\sigma^{\mu \nu} = \frac{i}{2}[\gamma^{\mu},\gamma^{\nu}]$, $\gamma^5 = i\gamma^0\gamma^1\gamma^2\gamma^3$, and $\varepsilon^{0123} = -1$. The scale parameter \( \mu \sim m_b \sim 5 \) GeV separates perturbative and non-perturbative contributions of the strong interaction. The perturbative contribution is contained in the Wilson coefficients \( C_{7}(\mu) \), \( C_{9V}(\mu) \), and \( C_{10A}(\mu) \). The non-perturbative contribution arises mainly in the computation of the matrix elements of the Hamiltonian (\ref{Effective_Hamiltonian}) between the initial and final hadronic states.

Within the SM, at $ \mu_0 = 5$ GeV, the following numerical values for the Wilson coefficients can be obtained: \( C_1(\mu_0) = 0.241 \), \( C_2(\mu_0) = -1.1 \), \( C_3(\mu_0) = -0.0104 \), \( C_4(\mu_0) = 0.02433 \), \( C_5(\mu_0) = -0.00706 \), \( C_6(\mu_0) = 0.0294 \), \( C_{7}(\mu_0) = 0.312 \), \( C_{9V}(\mu_0) = -4.21 \), and \( C_{10A}(\mu_0) = 4.41 \) \cite{Buchalla:1995vs, PhysRevD.52.186,Melikhov:1998ws}.

The matrix element of the axial current is determined via the decay constant $f_{B^0_q}$ \cite{FlavourLatticeAveragingGroupFLAG:2024oxs}:

\begin{equation}
\label{formfactorBll}
    \langle 0 | \bar{q}(0)\gamma^\mu \gamma^5 b(0)|B^0_{d,s}(M,P)\rangle = if_{B^0_{d,s}} P^\mu.
\end{equation}

\begin{table}[ht]
\centering
\begin{tabular}{ || l | l | l | l | l || }
\hline
  Decay & $\mathcal{B}^{(exp)}$ & $\mathcal{B}^{(free)}$ & $\mathcal{B}^{(Coulomb)}$ & $\Delta\mathcal{K}_\text{C}$ \\
  \hline
$B^0_{s}\rightarrow \mu^+\mu^- [10^{-9}]$ & $3.83^{+0.44}_{-0.41}$ & $3.66\pm 0.14$  & $3.75\pm 0.14$ & $2.3\%$ \\
 & & & & \\
$B^0\rightarrow \mu^+\mu^- [10^{-11}]$  &  $<19$ &  $1.03 \pm 0.05$ &  $1.05\pm0.05$ &  $2.3\%$ \\
 & \text{(95\% CL)} & & & \\
\hline
$B^0_s\rightarrow e^+e^- [10^{-14}]$ &  $<9.4\cdot 10^5$ &  $8.60\pm0.36$ &  $8.79\pm0.37$ &  $2.3\%$ \\
 & \text{(95\% CL)} & & & \\
$B^0\rightarrow e^+e^- [10^{-15}]$ &  $<3.0\cdot 10^6$ &  $2.41\pm0.13$ &  $2.47\pm0.13$ &  $2.3\%$ \\
 & \text{(95\% CL)} & & & \\
\hline
$B^0_s\rightarrow \tau^+\tau^-[10^{-7}]$ &  $<6.8\cdot10^{6}$ &  $6.94 \pm 1.88$  &  $7.11\pm1.93$ &  $2.4\%$ \\
 & \text{(95\% CL)} & & & \\
$B^0\rightarrow \tau^+\tau^-[10^{-8}]$ &  $<2.1\cdot10^{7}$ &  $1.99 \pm 0.22$  &  $2.04\pm0.23$ &  $2.4\%$ \\
 & \text{(95\% CL)} & & & \\
\hline
\end{tabular}
\caption{Partial decay widths for $B^0_{d,s}\rightarrow \ell^+\ell^-$. Comparison of experimental data \cite{CMS:2022mgd, ParticleDataGroup:2024cfk}, theoretical SM predictions without Coulomb interaction \cite{Beneke:2019slt, Huang:2023nli}, and predictions including this interaction. The relative Coulomb correction is given by $\Delta\mathcal{K}_C \equiv \mathcal{K}_C-1$. The full correction $\mathcal{K}_{\text{QED}}$ including soft photons is given in Table~\ref{tab:Bll_full_QED} in Appendix \ref{full_QED_corr}.}
\label{tab:branching_ratio_Bll}
\end{table}

Application of the Coulomb correction for the leptonic decay yields the following value for the partial decay width of $B^0_{d,s}\to \ell^+\ell^-$:
\begin{equation}
    \label{Bll_branching_Furry_method}
    \Gamma^{(\text{Coulomb})}_{B^0_{d,s}\to \ell^+\ell^-} = \Gamma^{(\text{free})}_{B^0_{d,s}\to \ell^+\ell^-}\cdot \mathcal{K}_\text{C}(v)
\end{equation}
where $\Gamma^{(\text{free})}_{B^0_{d,s}\to \ell^+\ell^-}$ is the decay width without accounting for the Coulomb interaction, the exact formula for which is given in \cite{Beneke:2019slt}. The argument $v$, which we substitute into the correction factor $\mathcal{K}_\text{C}(v)$, is equal to:
\begin{equation*}
    v = \frac{\sqrt{1-4(m_{\ell}/M_{B})^2}}{1-2(m_{\ell}/M_{B})^2},
\end{equation*}
and represents the relative velocity of the $\ell^+\ell^-$ pair.

A comparison of experimental data and theoretical predictions for the decays $B_{d,s}^0\rightarrow \ell^+\ell^-$ is presented in Table \ref{tab:branching_ratio_Bll}. As a measure of the theory-experiment discrepancy, we use the quantity:
\begin{equation}
    \label{rho_discrepancy}
    \delta = \frac{|\mathcal{B}^{(exp)} - \mathcal{B}^{(theory)}|}{\mathcal{B}^{(exp)}},
\end{equation}
where $\mathcal{B}^{(exp)}$ and $\mathcal{B}^{(theory)}$ are the experimental and theoretical branching fractions.

The SM predictions for the decays $B_{d,s}^0 \rightarrow \mu^+\mu^-$ without account of the Coulomb interaction are taken from \cite{Beneke:2019slt}. The value of the Coulomb correction for all decays is about $\Delta\mathcal{K}_C \equiv \mathcal{K}_C-1 = 2.3\%$. It can be seen that for the decay $B_s^0 \rightarrow \mu^+\mu^-$, accounting for the Coulomb interaction improves the agreement between the SM predictions and the experimental data: the discrepancy between the average experimental value and the corrected theoretical prediction reduces to $\delta = 2\%$. It should be noted that this improvement remains smaller than the current experimental and theoretical uncertainties. The experimental error is $11\%$ \cite{CMS:2022mgd}, while the SM prediction has a theoretical uncertainty of about $5\%$, coming from several sources: the decay constant $f_{B^0_{s}}$ \cite{FlavourLatticeAveragingGroupFLAG:2024oxs}, the CKM parameters, and other inputs \cite{Beneke:2019slt}.

For completeness, we have evaluated the full soft-photon QED correction (\ref{Isidori_factor}), which accounts for both the soft and Coulomb contributions. The values of $\mathcal{K}_{\text{QED}}$ for different photon energy cutoffs $\Delta E$ are listed in Table~\ref{tab:Bll_full_QED} in the appendix \ref{full_QED_corr}.

\section{Coulomb Interaction in \texorpdfstring{$B^0_{d,s}\to h^0\ell^+\ell^-$}{B0d,s → h0ℓ+ℓ-} Decays}
\label{sec4}

This section applies the Coulomb correction to analyse the decays $B^0\rightarrow h^0 \ell^+\ell^-$, where $h^0$ is a neutral pseudoscalar meson. The differential decay widths are given by:
\begin{equation}
    \frac{d\Gamma^{(\text{Coulomb})}_{B^0_{d,s}\to h^0\ell^+\ell^-}}{d\hat{s}} = \frac{d\Gamma^{(\text{free})}_{B^0_{d,s}\to h^0\ell^+\ell^-}}{d\hat{s}} \cdot\mathcal{K}_\text{C}(v),
    \label{differential_width_general}
\end{equation}
where $\hat{s} = q^2/M_B^2 = (p_B - p_h)^2/M_B^2$, $\hat{m} = (m_{\ell}/M_B)^2$, and $d\Gamma^{(\text{free})}_{B^0_{s, d}\to h^0\ell^+\ell^-}/d\hat{s}$ is the differential width without Coulomb interaction, whose exact form is provided in the \cite{PhysRevD.57.6814}. Relative velocity $v$ is equal to:
\begin{equation*}
    v = \frac{\sqrt{1-4\hat{m}/\hat{s}}}{1-2\hat{m}/\hat{s}}.
\end{equation*}


\begin{figure}[ht]
\centering
\includegraphics[width=80mm]{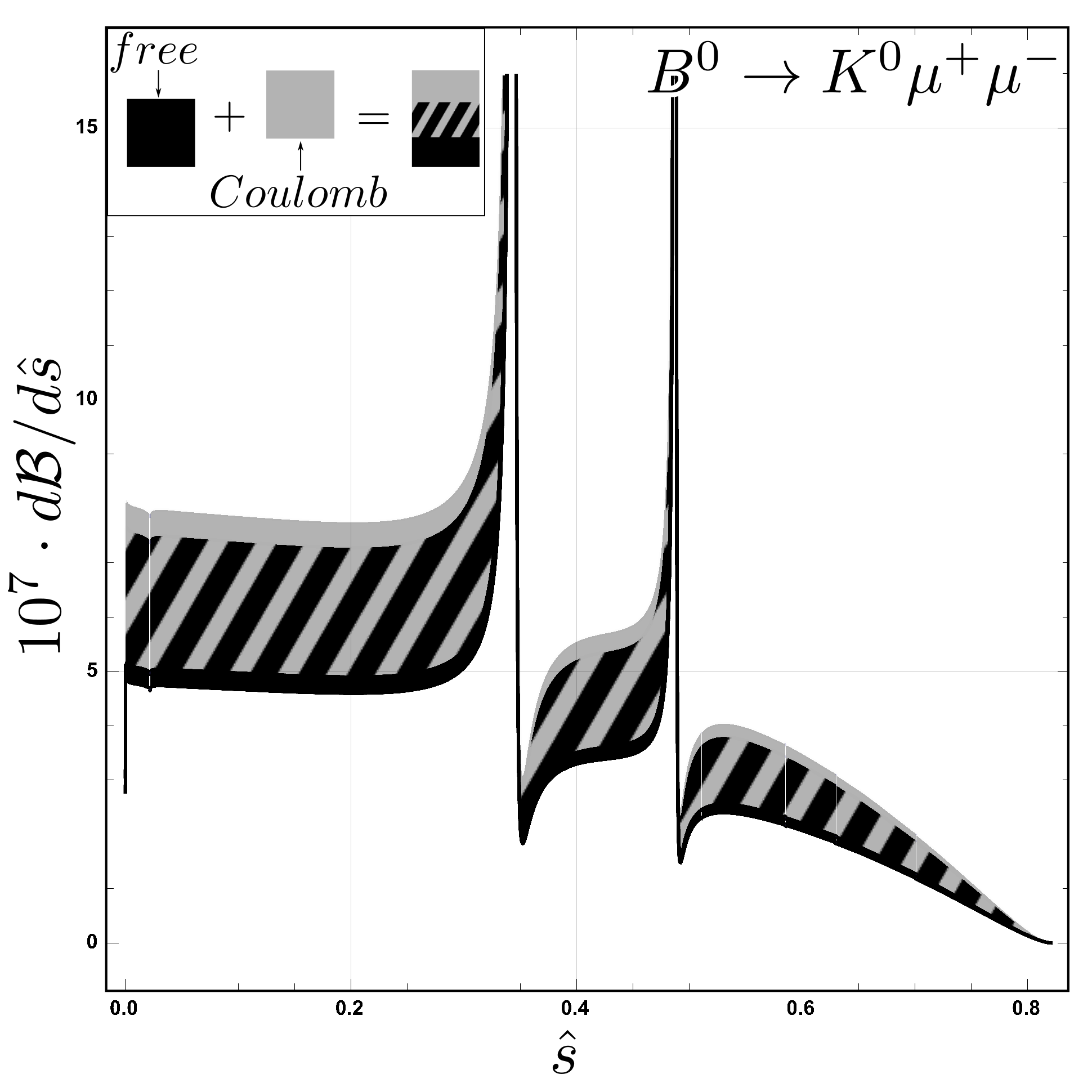}
\caption{Dependence of the differential branching fraction $10^7 d\mathcal{B}/d\hat{s}$ for the decay $B^0\rightarrow K^0 \mu^+\mu^-$ on $\hat{s} = (p_B-p_K)^2/M_{B}^2$ -- the squared transferred momentum normalized to the square of the $B$-meson mass. The black band corresponds to predictions without Coulomb interaction, the gray band -- with Coulomb interaction. The overlap region is indicated by black-gray hatching.}
\label{pic_Bkll}
\end{figure}

Within this study, calculations were performed for the decays $B^0\rightarrow \{K^0, \pi^0\} \ell^+\ell^-$ and $B_s^0\rightarrow \{\eta, \eta', K^0\} \ell^+\ell^-$. Differential, angular, and double differential distributions are shown in Figs. \ref{pic_Bhll}, \ref{pic_Bhll_angle} and \ref{pic_Bhll_3D} in Appendix \ref{app_plots}. A typical example of the differential distribution for the decay $B^0 \to K^0 \mu^+\mu^-$ is shown in Fig. \ref{pic_Bkll}. The main contribution to the calculation uncertainty is associated with the error in the hadronic transition form-factors (values taken from \cite{Melikhov:2000yu, Gubernari:2018wyi}). Accounting for the Coulomb interaction leads to a slight upward shift of the entire curve, remaining within the error band.

In accordance with the experimental procedure \cite{LHCb:2016ykl}, when calculating the total decay width, resonant peaks for $J/\psi$ and $\psi(2S)$ mesons are excluded in the regions:
\begin{equation}
    q^2 \in [8, 11]\text{GeV}^2, \text{ and } q^2 \in [12.5, 15]\text{GeV}^2.
    \label{constrains}
\end{equation}
To estimate the contribution of non-resonant processes in the excluded regions, the value of the differential width was assumed constant and equal to the arithmetic mean of the values at the boundaries of the intervals. The values of the partial decay widths are given in Table \ref{tab:branching_ratio_Bhll}.

The average Coulomb correction is defined as the ratio of the integrated widths:
\begin{equation}
    \label{mean_K_corr}
    \langle\mathcal{K}_C\rangle = \frac{\displaystyle\int\frac{d\Gamma^{(\text{Coulomb})}_{B^0_{d,s}\to h^0\ell^+\ell^-}}{d\hat{s}}\, d\hat{s}}{\displaystyle\int\frac{d\Gamma^{(\text{free})}_{B^0_{d,s}\to h^0\ell^+\ell^-}}{d\hat{s}}\, d\hat{s}}.
\end{equation}

\begin{table}[ht]
\centering
\begin{tabular}{ || l | l | l | l | l || }
\hline
  & $\mathcal{B}^{(exp)}$ & $\mathcal{B}^{(free)}$ & $\mathcal{B}^{(Coulomb)}$ & $\Delta\langle\mathcal{K}_C\rangle$\\ \hline
$B^0\rightarrow K^0e^+e^- [10^{-7}]$ & $2.5^{+1.1}_{-0.9}$ & $3.34\pm 0.37$& $3.42\pm 0.39$& $2.3 \%$\\
& & & & \\
$B^0\rightarrow K^0\mu^+\mu^-[10^{-7}]$ & $3.39 \pm$ & $3.33\pm 0.38$ & $3.41\pm 0.38$ & $2.3 \%$\\
& $ \pm 0.35$& & & \\
$B^0\rightarrow K^0\tau^+\tau^-[10^{-8}]$ & - & $5.0\pm 2.1$& $5.18\pm 2.2$& $3.3 \%$\\
& & & & \\ \hline
$B^0\rightarrow \pi^0e^+e^-[10^{-8}]$ & $ <8.4$ & $1.17\pm 0.3$ &  $1.19\pm 0.3$ & $2.3 \%$\\
& & & & \\
$B^0\rightarrow \pi^0\mu^+\mu^-[10^{-8}]$ & $< 6.9$ & $1.16\pm 0.3$ &  $1.19\pm 0.3$ & $2.3 \%$\\
& & & & \\
$B^0\rightarrow \pi^0\tau^+\tau^-[10^{-9}]$ & $-$  & $ 3.05\pm 0.73$ & $3.15\pm 0.76$& $3.0 \%$\\
& & & & \\ \hline
$B_s^0\rightarrow \eta e^+e^-[10^{-7}]$ & $-$ & $4.12\pm 0.79$ & $4.22\pm 0.80$ & $2.3 \%$\\
& & & & \\
$B_s^0\rightarrow \eta\mu^+\mu^-[10^{-7}]$ & $-$ & $4.11\pm0.79$ & $4.21\pm0.80$ & $2.4 \%$\\
& & & & \\
$B_s^0\rightarrow \eta\tau^+\tau^-[10^{-8}]$ & $-$ & $ 7.04\pm 1.2$ & $7.28\pm 1.3$& $3.3 \%$\\
& & & & \\ \hline
$B_s^0\rightarrow \eta' e^+e^-[10^{-7}]$ & $-$ & $3.04\pm 0.58$ & $3.11\pm 0.59$ & $2.3 \%$\\
& & & & \\
$B_s^0\rightarrow \eta'\mu^+\mu^-[10^{-7}]$ & $-$ & $3.03\pm 0.58$ & $3.10\pm 0.59$ & $2.3 \%$\\
& & & & \\
$B_s^0\rightarrow \eta'\tau^+\tau^-[10^{-8}]$ & $-$ & $ 2.37\pm 0.36$ & $2.47\pm0.38$& $\mathbf{3.9\%}$\\
& & & & \\ \hline
$B_s^0\rightarrow K^0e^+e^-[10^{-8}]$ & $-$ & $1.35\pm 0.34$ & $1.38\pm 0.35$ & $2.3 \%$\\
& & & & \\
$B_s^0\rightarrow K^0\mu^+\mu^-[10^{-8}]$ & $-$ & $1.34\pm0.34$ & $1.37\pm 0.35$ & $2.4 \%$\\
& & & & \\
$B_s^0\rightarrow K^0\tau^+\tau^-[10^{-9}]$ & $-$ & $ 2.55 \pm 0.59$ & $2.64\pm 0.62$& $3.3 \%$\\
& & & & \\ \hline
 \end{tabular}
 \caption{Partial decay widths for $B^0\rightarrow \{K^0, \pi^0\} \ell^+\ell^-$ and $B_s^0\rightarrow \{\eta, \eta', K^0\} \ell^+\ell^-$. Experimental data is taken from \cite{ParticleDataGroup:2024cfk}. The relative Coulomb correction is given by $\Delta\langle\mathcal{K}_C\rangle \equiv \langle\mathcal{K}_{\text{C}}\rangle-1$, where $\langle\mathcal{K}_{\text{C}}\rangle$ is defined in (\ref{mean_K_corr}). The full correction $\langle\mathcal{K}_{\text{QED}}\rangle$ including soft photons is given in Table~\ref{tab:Bhll_full_QED} in Appendix \ref{full_QED_corr}.}
    \label{tab:branching_ratio_Bhll}
\end{table}

For $B^0\to K^0\mu^+\mu^-$, the Coulomb corrections reduce the discrepancy between theory and experiment from $\delta = 1.7\%$ to $\delta = 0.5\%$. Given the 10\% uncertainties in both experimental and theoretical predictions, this level of agreement may reflect statistical fluctuations.

For the decay $B^0\to K^0e^+e^-$, the Coulomb correction formally increases the discrepancy from $\delta = 33\%$ to $\delta = 36\%$, while still remaining within the 45\% experimental uncertainty. 

For decays involving $\tau$-leptons, the Coulomb correction is of the order of $\Delta\langle\mathcal{K}_C\rangle \equiv \langle\mathcal{K}_{\text{C}}\rangle-1 \approx 4\%$. While this is currently smaller than the dominant form-factor uncertainties, it constitutes a systematic effect that must be accounted for in future high-precision experimental analyses.

As for the leptonic decays, we have obtained the full soft-photon QED correction for the semileptonic channels. The results are presented in Table~\ref{tab:Bhll_full_QED} of the appendix \ref{full_QED_corr}.

\section{Coulomb Interaction in \texorpdfstring{$B^0_{d,s}\to V^0\ell^+\ell^-$}{B0d,s → V0ℓ+ℓ-} Decays}
\label{sec5}

This section examines the Coulomb interaction in the decays $B^0\rightarrow \{K^{0*}, \rho^{0}\} \ell^+\ell^-$ and $B_s^0\rightarrow \{\phi, K^{0*}\} \ell^+\ell^-$ involving a neutral vector meson. The differential decay width is determined similarly to (\ref{differential_width_general}) with the pseudoscalar meson $h^0$ replaced by the vector meson $V^0$. The explicit expression for $d\Gamma^{(\text{free})}_{B^0_{s, d}\to V^0\ell^+\ell^-}/d\hat{s}$ is derived from the theoretical framework described in the \cite{PhysRevD.57.6814}.

Differential, angular and double differential distributions are shown on Figs. \ref{pic_BVll}, \ref{pic_BVll_angle} and \ref{pic_BVll_3D} in Appendix \ref{app_plots}. The predicted partial widths, obtained with the procedure of cutting out the peak regions defined in (\ref{constrains}), are presented in Table \ref{tab:branching_ratio_BVll}.

\begin{table}[ht]
\centering
\begin{tabular}{ || l | l | l | l | l || }
\hline 
  & $\mathcal{B}^{(exp)}$ & $\mathcal{B}^{(free)}$ & $\mathcal{B}^{(Coulomb)}$ & $\Delta\langle\mathcal{K}_C\rangle$\\ \hline
$B^0\rightarrow K^{0*}e^+e^- [10^{-6}]$& $1.19\pm$& $1.36\pm 0.15$& $1.40\pm 0.15$& $2.4 \%$\\ 
& $\pm  0.20$& & & \\
$B^0\rightarrow K^{0*}\mu^+\mu^-[10^{-6}]$& $1.06\pm$& $0.94\pm 0.11$& $1.02\pm 0.11$& $2.4\%$\\
& $\pm  0.09$& & & \\ 
$B^0\rightarrow K^{0*}\tau^+\tau^-[10^{-8}]$& $-$& $6.67\pm 0.87$& $6.92\pm 0.90$& $3.7 \%$\\ 
& & & & \\ \hline
$B^0\rightarrow \rho^{0} e^+e^-[10^{-6}]$& $-$& $1.07\pm 0.18$&  $1.10\pm 0.19$& $2.4 \%$\\ 
& & & & \\ 
$B^0\rightarrow \rho^{0}\mu^+\mu^-[10^{-7}]$& $-$& $8.9\pm 1.6$&  $9.1\pm 1.7$& $2.4 \%$\\
& & & & \\ 
$B^0\rightarrow \rho^{0}\tau^+\tau^-[10^{-7}]$& $-$  & $ 1.05\pm 0.19$& $ 1.09\pm 0.20$& $3.4\%$\\
& & & & \\ \hline
$B_s^0\rightarrow \phi e^+e^-[10^{-6}]$& $-$ & $1.52\pm 0.17$& $1.56\pm 0.17$& $2.4 \%$\\ 
& & & & \\
$B_s^0\rightarrow \phi\mu^+\mu^-[10^{-6}]$& $0.84\pm$& $1.15\pm0.14$& $1.18\pm0.15$& $2.4 \%$\\
& $\pm  0.4$& & & \\
$B_s^0\rightarrow \phi\tau^+\tau^-[10^{-8}]$& $-$ & $ 8.11\pm 0.10$& $ 8.41\pm 0.11$& $3.8\%$\\ 
& & & & \\ \hline
$B_s^0\rightarrow K^{0*} e^+e^-[10^{-8}]$& $-$ & $5.56\pm 0.65$& $5.69\pm 0.67$& $2.4 \%$\\ 
& & & & \\
$B_s^0\rightarrow K^{0*}\mu^+\mu^-[10^{-8}]$& $2.9\pm$& $4.42\pm 0.56$& $4.53\pm 0.57$& $2.4 \%$\\
& $\pm  1.1$& & & \\
$B_s^0\rightarrow K^{0*}\tau^+\tau^-[10^{-9}]$& $-$ & $ 4.30\pm 0.50$& $4.45\pm0.52$& $3.5\%$\\ 
& & & & \\ \hline
 \end{tabular}
 \caption{Partial decay widths for $B^0_{d,s}\rightarrow V^0\ell^+\ell^-$. Experimental data is taken from \cite{ParticleDataGroup:2024cfk}. The relative Coulomb correction is given by $\Delta\langle\mathcal{K}_C\rangle \equiv \langle\mathcal{K}_{\text{C}}\rangle-1$, where $\langle\mathcal{K}_{\text{C}}\rangle$ is defined similarly to (\ref{mean_K_corr}). The full correction $\langle\mathcal{K}_{\text{QED}}\rangle$ including soft photons is given in Table~\ref{tab:BVll_full_QED} in Appendix \ref{full_QED_corr}.}
    \label{tab:branching_ratio_BVll}
\end{table}

An interesting feature can be observed in the angular distributions $d\mathcal{B}/d\cos\theta$, where $\theta$ is the angle between the direction of the neutral hadron $h^0$ and the positive lepton $\ell^+$ in the $\ell^+\ell^-$ rest frame.  When comparing the $B^{0}_{d,s}\to V^0\mu^+\mu^-$ and $B^{0}_{d,s}\to V^0 e^+ e^-$ differential distributions, the curvature changes from convex to concave for some decay modes (this transition is present in  $B^{0}\to \rho^{0} \ell^+\ell^-$ and $B^{0}_s\to K^{0*}\ell^+\ell^-$, but absent in $B^{0}\to K^{0*}\ell^+\ell^-$ and  $B^{0}_s\to\phi\ell^+\ell^-$). This change in behavior is almost entirely determined by the energy region  $\hat{s} \in (4m_e^2/M_{B^{0}_{d,s}}^2; 4m_\mu^2/M_{B^{0}_{d,s}}^2)$. While the muonic decay is kinematically forbidden in this region, the electronic channel still yields a non-zero differential distribution. 

In the $B^0\rightarrow K^{0*}\mu^+\mu^-$ decay, accounting for Coulomb interaction reduces the discrepancy with experiment from $\delta = 11\%$ to $\delta = 4\%$. For the $B^0\rightarrow K^{0*}e^+e^-$ decay, the Coulomb correction increases the discrepancy from $\delta = 14\%$ to $\delta = 17\%$, though the prediction remains within experimental uncertainty. The decays $B^0_{d,s}\rightarrow V^0\ell^+\ell^-$ with $\tau$-leptons demonstrate a Coulomb correction up to $\Delta\langle\mathcal{K}_C\rangle \equiv \langle\mathcal{K}_{\text{C}}\rangle-1 = 3.8\%$.

The full soft-photon QED corrections for $B^0_{d,s}\to V^0\ell^+\ell^-$ are listed in Table~\ref{tab:BVll_full_QED} in the appendix \ref{full_QED_corr}.

\section{Coulomb Interaction in \texorpdfstring{$B^0_{d,s}\to \gamma\ell^+\ell^-$}{B0d,s → γℓ+ℓ-} Decays}
\label{sec6}

In this section, we provide the Coulomb interaction in the decays $B^0_{d,s}\rightarrow \gamma \ell^+\ell^-$. The differential decay width is determined similarly to (\ref{differential_width_general}) with the pseudoscalar meson $h^0$ replaced by the photon $\gamma$. The explicit expression for $d\Gamma^{(\text{free})}_{B^0_{s, d}\to \gamma\ell^+\ell^-}/d\hat{s}$ is taken from \cite{Kozachuk:2017mdk}.

Differential, angular and double differential distributions are shown on Figs. \ref{pic_Bgammall}, \ref{pic_Bgammall_angle} and \ref{pic_Bgammall_3D} in Appendix \ref{app_plots}.  

The partial decay widths for $B^0_{d,s}\to \gamma e^+e^-$ and $B^0_{d,s}\to \gamma \mu^+\mu^-$ are typically calculated within the interval $q^2 \in [1, 6] \text{ GeV}^2$. This choice is motivated by the relatively small contribution from charming loops in this region, which is at the level of a few percent. Consequently, the branching fractions can be predicted with controlled accuracy, limited primarily by the uncertainty in the form factors \cite{Kozachuk:2017mdk}.

In the considered decays, the resonant peaks $\rho^0$, $\omega$, $J/\psi$ and $\psi(2S)$ influence on the differential decay width. The $J/\psi$ and $\psi(2S)$ peaks are removed according to the procedure (\ref{constrains}), while the $\rho^0$ and $\omega$ fall outside integration region $q^2 \in [1, 6] \text{ GeV}^2$. Predictions for partial widths are presented in Table \ref{tab:branching_ratio_Bgammall}.

To compute decays involving $\tau$-leptons, a different procedure must be employed, as a large part of the interval $q^2 \in [1, 6] \text{ GeV}^2$ is kinematically forbidden. The only resonant peak that needs to be excluded is $\psi(2S)$, which we remove according to (\ref{constrains}). The partial width, presented in Table \ref{tab:branching_ratio_Bgammall}, is obtained by integrating over the entire allowed energy range $q^2 \in [4m_{\tau}^2, M_{B^0_{d,s}}^2]$.

\begin{table}[ht]
\centering
\begin{tabular}{ || l | l | l | l | l || }
\hline 
   $q^2\in [1 , 6] \text{ GeV}^2$& $\mathcal{B}^{(exp)}$ & $\mathcal{B}^{(free)} $& $\mathcal{B}^{(Coulomb)}$ & $\Delta\langle\mathcal{K}_C\rangle$ \\ \hline
$B_s^0\rightarrow \gamma e^+e^- [10^{-9}]$& $-$ & $2.97\pm$& $3.04\pm 0.50$& $2.4 \%$\\ 
& & $\pm0.49$& & \\
$B_s^0\rightarrow \gamma\mu^+\mu^-[10^{-9}]$& $<119$ & $3.03\pm$& $3.10\pm 0.50$& $2.4\%$\\
& $\text{ at }90\%\text{ CL}$& $\pm 0.50$& & \\  \hline
$B^0\rightarrow \gamma e^+e^-[10^{-12}]$& $-$& $5.09\pm $&  $5.21\pm 0.61$& $2.3 \%$\\ 
& & $\pm 0.59$& & \\ 
$B^0\rightarrow \gamma\mu^+\mu^-[10^{-12}]$& $-$& $5.31\pm$&  $5.43\pm 0.58$& $2.3 \%$\\
& & $\pm 0.57$& & \\  \hline\hline
$q^2\in [4m_{\tau}^2, M_{B_{d,s}}^2]$& $\mathcal{B}^{(exp)}$ & $\mathcal{B}^{(free)} $& $\mathcal{B}^{(Coulomb)}$ & $\Delta\langle\mathcal{K}_C\rangle$\\ \hline
$B_s^0\rightarrow \gamma \tau^+\tau^- [10^{-9}]$& $-$ & $1.13\pm$& $1.16\pm 0.31$& $2.7 \%$\\ 
& & $\pm 0.30$& & \\
$B^0\rightarrow \gamma\tau^+\tau^-[10^{-11}]$& $-$ & $3.10\pm$& $3.19\pm 0.95$& $2.7\%$\\
& & $\pm 0.93$& & \\  \hline
 \end{tabular}
 \caption{Partial decay widths for $B^0_{d,s}\rightarrow \gamma\ell^+\ell^-$ in the interval $q^2\in [1 , 6] \text{ GeV}^2$ for $\ell = \{e, \mu\}$ and $q^2\in [4m_{\tau}^2, M_{B_{d,s}}^2]$ for $\ell = \tau$. Experimental data is taken from  \cite{LHCb:2024uff}. The relative Coulomb correction is given by $\Delta\langle\mathcal{K}_C\rangle \equiv \langle\mathcal{K}_{\text{C}}\rangle-1$, where $\langle\mathcal{K}_{\text{C}}\rangle$ is defined similarly to (\ref{mean_K_corr}). The full correction $\langle\mathcal{K}_{\text{QED}}\rangle$ including soft photons is given in Table~\ref{tab:Bgammall_full_QED} in Appendix \ref{full_QED_corr}.}
    \label{tab:branching_ratio_Bgammall}
\end{table}

The full soft-photon QED corrections for radiative leptonic decays are given in Table~\ref{tab:Bgammall_full_QED} in the appendix \ref{full_QED_corr}.

\section{Conclusion}

In the present work:

\begin{itemize}
    \item We have systematically investigated the Coulomb correction in relativistic regime. In particular, for scalar particles we compared the non-relativistic Gamow-Sommerfeld-Sakharov (GSS) factor and the Crater-Alstine-Sazdjian (CAS) approach of exact relativistic two-particle equations. Using CAS equations, we derived the Coulomb correction (\ref{eq:CAS_factor}, Appendix \ref{CAS_derivation}). We then placed the Coulomb correction in the context of full QED soft-photon calculation. We recalled that the complete correction factorises as $\mathcal{K}_{\text{QED}} = \mathcal{K}_{\text{soft}} \cdot \mathcal{K}_{\text{C}}$ (up to negligible terms). The soft factor $\mathcal{K}_{\text{soft}}$ depends on the experimental photon-energy cutoff $\Delta E$ and is routinely simulated by Monte Carlo tools such as \textsc{photos}. The Coulomb factor $\mathcal{K}_{\text{C}}$, however, is not included in those simulations. Based on the conducted analysis, we conclude that the Coulomb correction, although often associated with threshold effects, remains non‑zero even for relativistic decays and should not be overlooked in precision theoretical predictions.

    \item The Coulomb correction has been obtained for:

    a. Ultra-rare leptonic decays $B^0_{d,s}\to \ell^+\ell^-$, see Table \ref{tab:branching_ratio_Bll};
    
    b. Rare semileptonic decays $B^0\rightarrow \{K^0, \pi^0\} \ell^+\ell^-$ and $B_s^0\rightarrow \{\eta, \eta', K^0\} \ell^+\ell^-$ involving pseudoscalar mesons, see Table \ref{tab:branching_ratio_Bhll}, Figs. \ref{pic_Bkll}, \ref{pic_Bhll}, \ref{pic_Bhll_angle}, \ref{pic_Bhll_3D};
    
    c. Rare semileptonic decays $B^0\rightarrow \{K^{0*}, \rho^0\} \ell^+\ell^-$ and $B_s^0\rightarrow \{\phi, K^{0*}\} \ell^+\ell^-$ involving vector mesons, see Table \ref{tab:branching_ratio_BVll}, Figs. \ref{pic_BVll}, \ref{pic_BVll_angle}, \ref{pic_BVll_3D};

    d. Rare radiative leptonic decays $B^0\rightarrow \gamma \ell^+\ell^-$, see Table \ref{tab:branching_ratio_Bgammall}, Figs. \ref{pic_Bgammall}, \ref{pic_Bgammall_angle}, \ref{pic_Bgammall_3D}.

    \item A complete analysis of the soft-photon QED corrections, including the factorised soft and Coulomb contributions, has been performed. The results are summarised in Tables~\ref{tab:Bll_full_QED} (leptonic), \ref{tab:Bhll_full_QED} and \ref{tab:BVll_full_QED} (semileptonic), and \ref{tab:Bgammall_full_QED} (radiative leptonic) in the appendix \ref{full_QED_corr}.

    \item The following key results have been obtained:

   a. Accounting for the Coulomb interaction in the decay $B_s^0 \to \mu^+\mu^-$ reduces the discrepancy between the average experimental value and the theoretical prediction within the SM to $\delta = 2\%$. However, this improvement is small compared to the current experimental (11\% \cite{CMS:2022mgd}) and theoretical (5\% \cite{FlavourLatticeAveragingGroupFLAG:2024oxs}) uncertainties for the partial width of this decay.

   b. In the decay $B^0\to K^{0*}\mu^+\mu^-$, the correction reduces the discrepancy between theoretical predictions and experimental data from $\delta = 11\%$ to $\delta = 4\%$, against the background of an 8\% experimental error \cite{ParticleDataGroup:2024cfk}.

   c. In the decay $B^0\rightarrow K^{0*}e^+e^-$, the Coulomb correction increases the discrepancy between theoretical predictions and experimental data from $\delta = 14\%$ to $\delta = 17\%$, yet remains within the experimental uncertainty (17\%, \cite{ParticleDataGroup:2024cfk}).
   
   d. In the decay $B^0\to K^0\mu^+\mu^-$, the correction reduces the discrepancy between theoretical predictions and experimental data from $\delta = 1.7\%$ to $\delta = 0.5\%$, against the background of a 10\% experimental error \cite{ParticleDataGroup:2024cfk}.

   e. In decays with $\tau$-leptons in the final state (particularly in $B_s^0\rightarrow \eta'\tau^+\tau^-$ and $B_s^0\rightarrow \phi\tau^+\tau^-$), the correction reaches $\mathcal{K}\sim 4\%$.

\end{itemize}

The obtained results demonstrate that accounting for Coulomb corrections is an essential element of the theoretical analysis of rare leptonic and semileptonic decays of neutral $B$-mesons. In particular, for decays with electrons and muons, the corrections (2-3\%), although smaller than the current experimental uncertainties (5-15\% \cite{ParticleDataGroup:2024cfk}), may shift the central values and potentially improve the agreement between theory and experiment. For channels with $\tau$-leptons, the corrections reach 4\%. This suggests that the discussed correction may become significant for future high-precision experiments in $B$-physics, especially those involving $\tau$-leptons in the final states.

\section*{Funding}

For one of the authors, N.V. Nikitin, this work was supported by Russian Science Foundation (grant no. 25-22-00614 rare fourlepton decays of heavy mesons in the orthogonal amplitude technique)

\section*{Conflict of interest}

The authors of this work declare that they have no conflicts of interest.

\section*{Acknowledgments}
We are deeply grateful to A.B. Arbuzov, S.P. Baranov, and D.I. Melikhov for valuable comments and helpful discussions of this work.

\bibliography{biblio}

@article{LHCb:2024onj,
  author = "R. Aaij and others",
  title = "",
  collaboration = "LHCb",
  journal = "JHEP",
  volume = "09",
  pages = "026",
  year = "2024",
  doi = "10.1007/JHEP09(2024)026"
}

@article{LHCb:2013tgx,
    author = "Aaij, R and others",
    collaboration = "LHCb",
    title = "{Differential branching fraction and angular analysis of the decay $B_s^0\to\phi\mu^{+}\mu^{-}$}",
    eprint = "1305.2168",
    archivePrefix = "arXiv",
    primaryClass = "hep-ex",
    reportNumber = "CERN-PH-EP-2013-078, LHCB-PAPER-2013-017",
    doi = "10.1007/JHEP07(2013)084",
    journal = "JHEP",
    volume = "07",
    pages = "084",
    year = "2013"
}

@article{LHCb:2024uff,
  author = "R. Aaij and others",
  title = "",
  collaboration = "LHCb",
  journal = "JHEP",
  volume = "07",
  pages = "101",
  year = "2024",
  doi = "10.1007/JHEP07(2024)101"
}

@article{Kozachuk:2017mdk,
    author = "Kozachuk, Anastasiia and Melikhov, Dmitri and Nikitin, Nikolai",
    title = "{Rare FCNC radiative leptonic $B_{s,d}\to \gamma l^+l^-$ decays in the standard model}",
    eprint = "1712.07926",
    archivePrefix = "arXiv",
    primaryClass = "hep-ph",
    doi = "10.1103/PhysRevD.97.053007",
    journal = "Phys. Rev. D",
    volume = "97",
    number = "5",
    pages = "053007",
    year = "2018"
}

@article{LHCb:2022qnv,
  author = "R. Aaij and others",
  title = "",
  collaboration = "LHCb",
  journal = "Phys. Rev. Lett.",
  volume = "131",
  number = "5",
  pages = "051803",
  year = "2023",
  doi = "10.1103/PhysRevLett.131.051803"
}

@article{LHCb:2021trn,
  author = "R. Aaij and others",
  title = "",
  collaboration = "LHCb",
  journal = "Nature Phys.",
  volume = "18",
  number = "3",
  pages = "277-282",
  year = "2022",
  doi = "10.1038/s41567-023-02095-3"
}

@article{PhysRevLett.128.041801,
  author = "R. Aaij and others",
  title = "",
  collaboration = "LHCb",
  journal = "Phys. Rev. Lett.",
  volume = "128",
  number = "4",
  pages = "041801",
  year = "2022",
  doi = "10.1103/PhysRevLett.128.041801"
}

@article{LHCb:2021awg,
  author = "R. Aaij and others",
  title = "",
  collaboration = "LHCb",
  journal = "Phys. Rev. D",
  volume = "105",
  number = "1",
  pages = "012010",
  year = "2022",
  doi = "10.1103/PhysRevD.105.012010"
}

@article{LHCb:2017rmj,
  author = "R. Aaij and others",
  title = "",
  collaboration = "LHCb",
  journal = "Phys. Rev. Lett.",
  volume = "118",
  number = "19",
  pages = "191801",
  year = "2017",
  doi = "10.1103/PhysRevLett.118.191801"
}

@article{LHCb:2016ykl,
  author = "R. Aaij and others",
  title = "",
  collaboration = "LHCb",
  journal = "JHEP",
  volume = "04",
  pages = "142",
  year = "2017",
  doi = "10.1007/JHEP11(2016)047"
}

@article{LHCb:2014cxe,
  author = "R. Aaij and others",
  title = "",
  collaboration = "LHCb",
  journal = "JHEP",
  volume = "06",
  pages = "133",
  year = "2014",
  doi = "10.1007/JHEP06(2014)133"
}

@article{CMS:2024atz,
  author = "{A. Hayrapetyan} and others",
  title = "",
  collaboration = "CMS",
  journal = "Phys. Lett. B",
  volume = "864",
  pages = "139406",
  year = "2025",
  doi = "10.1016/j.physletb.2025.139406"
}

@article{CMS:2024syx,
  author = "{A. Hayrapetyan} and others",
  title = "",
  collaboration = "CMS",
  journal = "Rept. Prog. Phys.",
  volume = "87",
  number = "7",
  pages = "077802",
  year = "2024",
  doi = "10.1088/1361-6633/ad4e65"
}

@article{CMS:2022mgd,
  author = "{A. Tumasyan} and others",
  title = "",
  collaboration = "CMS",
  journal = "Phys. Lett. B",
  volume = "842",
  pages = "137955",
  year = "2023",
  doi = "10.1016/j.physletb.2023.137955"
}

@article{ATLAS:2018cur,
  author = "{M. Aaboud} and others",
  title = "",
  collaboration = "ATLAS",
  journal = "JHEP",
  volume = "04",
  pages = "098",
  year = "2019",
  doi = "10.1007/JHEP04(2019)098"
}

@article{Belle:2024cis,
  author = "{I. Adachi} and others",
  title = "",
  collaboration = "Belle and Belle-II",
  journal = "Phys. Rev. Lett.",
  volume = "133",
  number = "10",
  pages = "101804",
  year = "2024",
  doi = "10.1103/PhysRevLett.133.101804"
}

@article{PhysRevLett.131.051804,
  author = "{L. Aggarwal} and others",
  title = "",
  collaboration = "Belle-II",
  journal = "Phys. Rev. Lett.",
  volume = "131",
  number = "5",
  pages = "051804",
  year = "2023",
  doi = "10.1103/PhysRevLett.131.051804"
}

@article{Belle-II:2022fky,
  author = "{F. Abudin} and others",
  title = "",
  collaboration = "Belle-II",
  eprint = "2206.05946",
  archivePrefix = "arXiv",
  primaryClass = "hep-ex",
  year = "2022"
}

@article{Buras:1995iy,
  author = "A. J. Buras",
  title = "",
  journal = "Nucl. Instrum. Meth. A",
  volume = "368",
  pages = "1-20",
  year = "1995",
  doi = "10.1016/0168-9002(95)00869-1"
}

@article{Beneke:2019slt,
  author = "M. Beneke and C. Bobeth and R. Szafron",
  title = "",
  journal = "JHEP",
  volume = "10",
  pages = "232",
  year = "2019",
  doi = "10.1007/JHEP10(2019)232"
}

@article{PhysRevD.57.6814,
  author = "D. Melikhov and N. Nikitin and S. Simula",
  title = "",
  journal = "Phys. Rev. D",
  volume = "57",
  pages = "6814",
  year = "1998",
  doi = "10.1103/PhysRevD.57.6814"
}

@article{PhysRevD.70.114028,
  author = "D. Melikhov and N. Nikitin",
  title = "",
  journal = "Phys. Rev. D",
  volume = "70",
  pages = "114028",
  year = "2004",
  doi = "10.1103/PhysRevD.70.114028"
}

@article{PhysRevD.101.096007,
  author = "A. Danilina and N. Nikitin and K. Toms",
  title = "",
  journal = "Phys. Rev. D",
  volume = "101",
  pages = "096007",
  year = "2020",
  doi = "10.1103/PhysRevD.101.096007"
}

@article{Gamow1928,
  author = "G. Gamow",
  title = "",
  journal = "Zeitschrift f{\"u}r Physik",
  volume = "51",
  number = "3",
  pages = "204-212",
  year = "1928",
  doi = "10.1007/BF01343196"
}

@book{sommerfeld1921atombau,
  author = "A. Sommerfeld",
  title = "Atombau und Spektrallinien",
  publisher = "F. Vieweg \& Sohn",
  year = "1921",
}

@article{Sakharov:1991,
  author = "A. D. Sakharov",
  title = "",
  journal = "Sov. Phys. Usp.",
  volume = "34",
  number = "5",
  pages = "375–377",
  year = "1991",
}

@article{Crater:1983ew,
  author = "H. W. Crater and P. Van Alstine",
  title = "",
  journal = "Annals Phys.",
  volume = "148",
  pages = "57-94",
  year = "1983",
  doi = "10.1016/0003-4916(83)90330-5"
}

@article{Crater:1994rt,
  author = "H. W. Crater and P. Van Alstine",
  title = "",
  journal = "Found. Phys.",
  volume = "24",
  pages = "297-328",
  year = "1994",
  doi = "10.1007/BF02313126"
}

@article{Sazdjian:1985pg,
  author = "H. Sazdjian",
  title = "",
  journal = "Phys. Rev. D",
  volume = "33",
  pages = "3401",
  year = "1986",
  doi = "10.1103/PhysRevD.33.3401"
}

@book{greiner,
  author = "W. Greiner",
  title = "Relativistic Quantum Mechanics: Wave Equations. 3rd Edition",
  publisher = "Springer-Verlag, New York",
  year = "2000",
  doi = "10.1007/978-3-662-04275-5"
}

@article{Buchalla:1995vs,
  author = "G. Buchalla and A. J. Buras and M. E. Lautenbacher",
  title = "",
  journal = "Rev. Mod. Phys.",
  volume = "68",
  pages = "1125-1144",
  year = "1996",
  doi = "10.1103/RevModPhys.68.1125"
}

@article{PhysRevD.52.186,
  author = "A. J. Buras and M. M{\"u}nz",
  title = "",
  journal = "Phys. Rev. D",
  volume = "52",
  pages = "186--195",
  year = "1995",
  doi = "10.1103/PhysRevD.52.186"
}

@article{ParticleDataGroup:2024cfk,
  author = "{S. Navas} and others",
  title = "",
  collaboration = "Particle Data Group",
  journal = "Phys. Rev. D",
  volume = "110",
  number = "3",
  pages = "030001",
  year = "2024",
  doi = "10.1103/PhysRevD.110.030001"
}

@article{Melikhov:1998ws,
  author = "D. Melikhov and N. Nikitin and S. Simula",
  title = "",
  journal = "Phys. Lett. B",
  volume = "430",
  pages = "332-340",
  year = "1998",
  doi = "10.1016/S0370-2693(98)00524-3"
}

@article{FlavourLatticeAveragingGroupFLAG:2024oxs,
  author = "{Y. Aoki} and others",
  title = "",
  collaboration = "Flavour Lattice Averaging Group (FLAG)",
  eprint = "2411.04268",
  archivePrefix = "arXiv",
  primaryClass = "hep-lat",
  year = "2024"
}

@article{Melikhov:2000yu,
  author = "D. Melikhov and B. Stech",
  title = "",
  journal = "Phys. Rev. D",
  volume = "62",
  pages = "014006",
  year = "2000",
  doi = "10.1103/PhysRevD.62.014006"
}

@article{Gubernari:2018wyi,
  author = "N. Gubernari and A. Kokulu and D. van Dyk",
  title = "",
  journal = "JHEP",
  volume = "01",
  pages = "150",
  year = "2019",
  doi = "10.1007/JHEP01(2019)150"
}

@article{LHCb:2021zwz,
    author = "Aaij, R. and others",
    title = "",
    collaboration = "LHCb",
    eprint = "2105.14007",
    archivePrefix = "arXiv",
    primaryClass = "hep-ex",
    reportNumber = "LHCb-PAPER-2021-014, CERN-EP-2021-092",
    doi = "10.1103/PhysRevLett.127.151801",
    journal = "Phys. Rev. Lett.",
    volume = "127",
    number = "15",
    pages = "151801",
    year = "2021"
}

@article{Buras:2001pn,
    author = "Buras, Andrzej J.",
    title = "",
    editor = "Zichichi, A.",
    eprint = "hep-ph/0101336",
    archivePrefix = "arXiv",
    reportNumber = "TUM-HEP-402-01",
    doi = "10.1142/9789812778253_0005",
    journal = "Subnucl. Ser.",
    volume = "38",
    pages = "200--337",
    year = "2002"
}

@article{Isidori:2007zt,
    author = "Isidori, Gino",
    title = "{Soft-photon corrections in multi-body meson decays}",
    eprint = "0709.2439",
    archivePrefix = "arXiv",
    primaryClass = "hep-ph",
    doi = "10.1140/epjc/s10052-007-0487-0",
    journal = "Eur. Phys. J. C",
    volume = "53",
    pages = "567--571",
    year = "2008"
}

@article{Arbuzov:2011ff,
    author = "Arbuzov, Andrej B. and Kopylova, Tatiana V.",
    title = "{On relativization of the Sommerfeld-Gamow-Sakharov factor}",
    eprint = "1111.4308",
    archivePrefix = "arXiv",
    primaryClass = "hep-ph",
    doi = "10.1007/JHEP04(2012)009",
    journal = "JHEP",
    volume = "04",
    pages = "009",
    year = "2012"
}

@article{PhysRevD.56.7276,
  title = {Two-loop corrections to the electromagnetic vertex for energies close to threshold},
  author = {Hoang, A. H.},
  journal = {Phys. Rev. D},
  volume = {56},
  issue = {11},
  pages = {7276--7283},
  numpages = {0},
  year = {1997},
  month = {Dec},
  publisher = {American Physical Society},
  doi = {10.1103/PhysRevD.56.7276}
}

@article{Solovtsova:2009zq,
    author = "Solovtsova, O. P. and Chernichenko, Yu. D.",
    title = "{Threshold resummation $S$-factor in QCD: The Case of unequal masses}",
    eprint = "0904.0754",
    archivePrefix = "arXiv",
    primaryClass = "hep-ph",
    doi = "10.1134/S1063778810090139",
    journal = "Phys. Atom. Nucl.",
    volume = "73",
    pages = "1612--1621",
    year = "2010"
}

@article{Arbuzov:1993qc,
    author = "Arbuzov, A. B.",
    title = "{On a novel equal time relativistic quasipotential equation for two scalar particles}",
    reportNumber = "JINR-E4-93-176",
    doi = "10.1007/BF02775767",
    journal = "Nuovo Cim. A",
    volume = "107",
    pages = "1263--1274",
    year = "1994"
}

@article{PhysRevD.3.2351,
  title = {Quasipotential Equation Corresponding to the Relativistic Eikonal Approximation},
  author = {Todorov, I. T.},
  journal = {Phys. Rev. D},
  volume = {3},
  issue = {10},
  pages = {2351--2356},
  numpages = {0},
  year = {1971},
  month = {May},
  publisher = {American Physical Society},
  doi = {10.1103/PhysRevD.3.2351},
  url = {https://link.aps.org/doi/10.1103/PhysRevD.3.2351}
}

@article{Yoon:2004sr,
    author = "Yoon, Jin-Hee and Wong, Cheuk-Yin",
    title = "{Relativistic generalization of the Gamow factor for fermion pair production or annihilation}",
    eprint = "nucl-th/0412019",
    archivePrefix = "arXiv",
    doi = "10.1088/0954-3899/31/2/007",
    journal = "J. Phys. G",
    volume = "31",
    pages = "149",
    year = "2005"
}

@article{PhysRevLett.120.011801,
  title = {Enhanced Electromagnetic Corrections to the Rare Decay ${B}_{s,d}\ensuremath{\rightarrow}{\ensuremath{\mu}}^{+}{\ensuremath{\mu}}^{\ensuremath{-}}$},
  author = {Beneke, Martin and Bobeth, Christoph and Szafron, Robert},
  journal = {Phys. Rev. Lett.},
  volume = {120},
  issue = {1},
  pages = {011801},
  numpages = {5},
  year = {2018},
  month = {Jan},
  publisher = {American Physical Society},
  doi = {10.1103/PhysRevLett.120.011801}
}

@article{Barberio:1993qi,
    author = "Barberio, Elisabetta and Was, Zbigniew",
    title = "{PHOTOS: A Universal Monte Carlo for QED radiative corrections. Version 2.0}",
    reportNumber = "CERN-TH-7033-93",
    doi = "10.1016/0010-4655(94)90074-4",
    journal = "Comput. Phys. Commun.",
    volume = "79",
    pages = "291--308",
    year = "1994"
}

@article{Golonka:2005pn,
    author = "Golonka, Piotr and Was, Zbigniew",
    title = "{PHOTOS Monte Carlo: A Precision tool for QED corrections in $Z$ and $W$ decays}",
    eprint = "hep-ph/0506026",
    archivePrefix = "arXiv",
    reportNumber = "IFJPAN-V-05-01, CERN-PH-TH-2005-091",
    doi = "10.1140/epjc/s2005-02396-4",
    journal = "Eur. Phys. J. C",
    volume = "45",
    pages = "97--107",
    year = "2006"
}

@article{Cali:2019nwp,
    author = "Cal{\'\i}, Stefano and Klaver, Suzanne and Rotondo, Marcello and Sciascia, Barbara",
    title = "{Impacts of radiative corrections on measurements of lepton flavour universality in $B \to D \ell \nu_{\ell}$ decays}",
    eprint = "1905.02702",
    archivePrefix = "arXiv",
    primaryClass = "hep-ph",
    doi = "10.1140/epjc/s10052-019-7254-x",
    journal = "Eur. Phys. J. C",
    volume = "79",
    number = "9",
    pages = "744",
    year = "2019"
}

@article{Huang:2023nli,
    author = "Huang, Yong-Kang and Shen, Yue-Long and Zhao, Xue-Chen and Zhou, Si-Hong",
    title = "{Complete analysis on QED corrections to B$_{q}$ {\textrightarrow} {\ensuremath{\tau}}$^{+}${\ensuremath{\tau}}$^{−}$}",
    eprint = "2301.00697",
    archivePrefix = "arXiv",
    primaryClass = "hep-ph",
    doi = "10.1007/JHEP10(2023)073",
    journal = "JHEP",
    volume = "10",
    pages = "073",
    year = "2023"
}

@article{Isidori:2020acz,
    author = "Isidori, Gino and Nabeebaccus, Saad and Zwicky, Roman",
    title = "{QED corrections in $ \overline{B}\to \overline{K}{\mathrm{\ell}}^{+}{\mathrm{\ell}}^{-} $ at the double-differential level}",
    eprint = "2009.00929",
    archivePrefix = "arXiv",
    primaryClass = "hep-ph",
    reportNumber = "CP3-Origins-2020-07, DNRF90, ZU-TH 30/20",
    doi = "10.1007/JHEP12(2020)104",
    journal = "JHEP",
    volume = "12",
    pages = "104",
    year = "2020"
}

@article{PhysRevD.108.L031502,
  title = {Structure-dependent QED effects in exclusive $B$ decays at subleading power},
  author = {Cornella, Claudia and K\"onig, Matthias and Neubert, Matthias},
  journal = {Phys. Rev. D},
  volume = {108},
  issue = {3},
  pages = {L031502},
  numpages = {6},
  year = {2023},
  month = {Aug},
  publisher = {American Physical Society},
  doi = {10.1103/PhysRevD.108.L031502}
}

@article{Buras:2012ru,
    author = "Buras, Andrzej J. and Girrbach, Jennifer and Guadagnoli, Diego and Isidori, Gino",
    title = "{On the Standard Model prediction for BR(B{s,d} to mu+ mu-)}",
    eprint = "1208.0934",
    archivePrefix = "arXiv",
    primaryClass = "hep-ph",
    reportNumber = "FLAVOUR(267104)-ERC-20, LAPTH-032-12, CERN-PH-TH-2012-210",
    doi = "10.1140/epjc/s10052-012-2172-1",
    journal = "Eur. Phys. J. C",
    volume = "72",
    pages = "2172",
    year = "2012"
}

@article{Bigi:2023cbv,
    author = "Bigi, Dante and Bordone, Marzia and Gambino, Paolo and Haisch, Ulrich and Piccione, Andrea",
    title = "{QED effects in inclusive semi-leptonic B decays}",
    eprint = "2309.02849",
    archivePrefix = "arXiv",
    primaryClass = "hep-ph",
    reportNumber = "CERN-TH-2023-145, MPP-2023-186",
    doi = "10.1007/JHEP11(2023)163",
    journal = "JHEP",
    volume = "11",
    pages = "163",
    year = "2023",
    note = "[Erratum: JHEP 03, 078 (2025)]"
}

@article{Isidori:2022bzw,
    author = "Isidori, Gino and Lancierini, Davide and Nabeebaccus, Saad and Zwicky, Roman",
    title = "{QED in $ \overline{B} ${\textrightarrow}$ \overline{K} ${\ensuremath{\ell}}$^{+}${\ensuremath{\ell}}$^{−}$ LFU ratios: theory versus experiment, a Monte Carlo study}",
    eprint = "2205.08635",
    archivePrefix = "arXiv",
    primaryClass = "hep-ph",
    doi = "10.1007/JHEP10(2022)146",
    journal = "JHEP",
    volume = "10",
    pages = "146",
    year = "2022"
}

\appendix
\onecolumngrid
\section{Relativistic two-particle equations}
\label{app1}

This appendix outlines the relativistic two-particle formalism developed by Crater, Alstine, and Sazdjian (CAS)\cite{Crater:1983ew,Crater:1994rt,Sazdjian:1985pg} based on relativistic classical constrained dynamics with the following quantisation. The CAS approach provides a rigorous, fully relativistic description of two-body interactions. Since the original derivation is rather involved we present here a condensed version of the formalism, focusing only on the essential steps required to derive the relativistic Coulomb factor used in our analysis.

Let us consider a system of two interacting scalar particles. In most general form the equations these scalar particles must obey can be written as follows:
\begin{equation}
\mathcal{H}_1 |\psi\rangle = \left(p_1^2 - m_1^2 - \Phi_1\right)|\psi\rangle = 0;\quad\mathcal{H}_2 |\psi\rangle = \left(p_2^2 - m_2^2 - \Phi_2\right)|\psi\rangle = 0
\end{equation}
where $p_i^2 = p_i^\mu p_{i\mu}$ is the squared momentum operator of the $i$-th particle, $m_i$ is the mass of the $i$-th particle, and $\Phi_i$ is the interaction operator. 

The key condition (or constraint) for these equations is the requirement $[\mathcal{H}_1,\mathcal{H}_2]|\psi\rangle=0$. In the simplest case, this can be achieved when:
\begin{equation}
\label{Newton_third_law}
    \Phi_1 = \Phi_2 \equiv \Phi(x_\perp)
\end{equation}
where $x_\perp^\mu = x^\mu - (x\cdot P)P^\mu/P^2$ is the transverse relative coordinate orthogonal to the total 4-momentum of the system $P^\mu = p_1^\mu + p_2^\mu = (E, \mathbf{P})$. Equation (\ref{Newton_third_law}) can be regarded as a relativistic generalization of Newton's third law.

We introduce the center of mass energy variables:
\begin{equation}
\epsilon_1 = \frac{P^2 + m_1^2 - m_2^2}{2\sqrt{P^2}}, \quad \epsilon_2 = \frac{P^2 + m_2^2 - m_1^2}{2\sqrt{P^2}}
\end{equation}
and the relative 4-momentum $q^\mu$:
\begin{equation}
p_1^\mu = \frac{\epsilon_1}{\sqrt{P^2}}P^\mu + q^\mu, \quad p_2^\mu = \frac{\epsilon_2}{\sqrt{P^2}}P^\mu - q^\mu
\end{equation}
Then the equation of motion takes the form:
\begin{equation}
\begin{split}
    & {\mathcal{H}}|\psi\rangle =  {\mathcal{H}}_1|\psi\rangle =  {\mathcal{H}}_2|\psi\rangle =  \big(\epsilon_1^2-m_1^2+  q^2- \Phi(x_\perp)\big)|\psi\rangle = \\
    & \big(\epsilon_2^2-m_2^2+  q^2- \Phi(x_\perp)\big)|\psi\rangle = (b^2(P^2,m_1^2,m_2^2) +   q^2 -  \Phi(x_\perp))|\psi\rangle = 0,
\end{split}
\end{equation}
where the relativistic invariant $b^2$ is defined as
\begin{equation}
\begin{split}
    &b^2(P^2,m_1^2,m_2^2) = \epsilon_1^2-m_1^2 = \epsilon_2^2-m_2^2 = \frac{1}{4P^2}(P^4-2P^2(m_1^2+m_2^2)+(m_1^2-m_2^2)^2)
\end{split}
\end{equation}
Using Todorov's kinematic variables \cite{PhysRevD.3.2351}:
\begin{equation}
m_s = \frac{m_1m_2}{\sqrt{s}}, \quad \epsilon_s = \frac{s - m_1^2 - m_2^2}{2\sqrt{s}}, \quad s \equiv P^2
\end{equation}
and an auxiliary momentum:
\begin{equation}
\mathcal{P}^\mu = q^\mu + \epsilon_s \frac{P^\mu}{\sqrt{s}}
\end{equation}
the equation takes the form:
\begin{equation}
\begin{split}
    & (b^2(s^2,m_1^2,m_2^2) +   q^2 -  \Phi(x_\perp))|\psi\rangle = (\epsilon_{s}^2 - m_s^2 +   q^2 -  \Phi(x_\perp))|\psi\rangle = (\mathcal{P}^2 - m_s^2 -  \Phi(x_\perp))|\psi\rangle = 0 
\end{split}
\end{equation}
This is the equation of relative motion for scalar particles in its most general form. Next, we need to specify the form of the interaction potential $\Phi(x_\perp)$. We are interested in the Coulomb interaction, which can be derived using the standard procedure:

\begin{equation}
    \mathcal{P}^\mu \to \mathcal{P}^\mu - A^\mu, \quad 
    A^\mu = (U(r), \mathbf{0}) = (-\alpha_{\text{em}}/r, \mathbf{0})
\end{equation}

which leads to the final equation in the center of mass system ($\mathbf{P} = 0$):

\begin{equation}
\label{CAS_eq_main}
    \left[ (\epsilon_s - U(r))^2 - \mathbf{\hat{q}}^2 - m_s^2 \right] \psi(\mathbf{r}) = 0
\end{equation}

where $\mathbf{q}^2 \equiv -\nabla^2$.

\section{Derivation of the Coulomb correction within the CAS approach.}
\label{CAS_derivation}

This appendix presents the derivation of the Coulomb correction factor (\ref{eq:CAS_factor}) within the CAS approach for the scalar decay \( B^0 \to S^+ S^- \). We begin by solving the CAS equation for the relative motion of the charged scalar \(S^+ S^-\)-pair in the s-wave (\(l=0\)) state. The obtained solution exhibits a weak divergence at the origin. We analyze this divergence, show that it lies outside the domain of validity of the CAS approximation, and estimate the range of particle velocities where its contribution is negligible. This allows us to arrive at the final expression for the correction factor.

To solve the CAS equation (\ref{CAS_eq_main}), we separate the variables in the wave function, $\psi ^{(Coulomb)}(\mathbf{r}) = R_{nl}(r)Y_{lm}(\theta, \phi)$, which yields the equation for the radial component:
\begin{equation}
\label{CAS_eq_radial}
\Bigg(\frac{d^2}{dz^2} + \frac{2}{z}\frac{d}{dz} - \frac{l(l+1)}{z^2} + \frac{2\eta}{z} + \frac{\alpha_{\text{em}}^2}{z^2} + 1 \Bigg)R_{nl}(r) = 0
\end{equation}
where $z = pr$, $p = \sqrt{\epsilon_s^2-m_s^2}$, $\eta = \alpha_{\text{em}}/v$, $v = p/\epsilon_s$.

Note that the velocity $v$ in this case coincides with the relative velocity:
\begin{equation}
\label{v_rel}
v = \frac{\sqrt{\epsilon_s^2 - m_s^2}}{m_s} = \frac{\sqrt{s^2 - 4m^2s}}{s - 2m^2} = v_{rel}
\end{equation}
The positive-frequency solution of equation (\ref{CAS_eq_radial}) is \cite{greiner}:
\begin{equation}
\label{Furry_scalar_solution}
R_{nl}^{(+)}(r) = \frac{|\Gamma(a)|}{\Gamma(b)} e^{\pi \eta/2} \cdot (2ipr)^{\mu - \frac{1}{2}}e^{ipr} F(a,b,-2ipr)
\end{equation}
\begin{equation}
\label{mu_definiton}
a = \mu + \frac{1}{2} + i\eta,\quad b = 2\mu +1;\quad \mu = \sqrt{\Big(l+\frac{1}{2}\Big)^2 - \alpha_{\text{em}}^2}
\end{equation}
where $\Gamma(z)$ is Euler's gamma function, and $F(a,b,z)$ is Kummer's confluent hypergeometric function.
The Coulomb correction is typically computed as the ratio $|\psi^{(\text{Coulomb})}(0)|^2 / |\psi^{(\text{free})}(0)|^2 = |\psi^{(\text{Coulomb})}(0)|^2$, where we have used that $|\psi^{(\text{free})}(\mathbf{r})|^2 = |e^{i\mathbf{p}\mathbf{r}}|^2 = 1$. It is important to note that in this standard approach, the wave function is evaluated not at the strict mathematical origin, but at a ``physical zero''—a distance much smaller than any typical scale of the problem and within the domain of validity of the underlying equations.

In our case, the decay $B \to S^+ S^-$ has zero orbital angular momentum ($l=0$), and therefore the wave function exhibits a weak divergence at $r \to 0$, associated with the factor $(2ipr)^{\mu-1/2}$. This divergence is removed by evaluating the wave function at a point shifted by a small distance $R$ from the center. Such a regularization is possible because the derivation of the CAS equations assumed a constant electromagnetic coupling, $\alpha_{\text{em}}(r) \approx \alpha_{\text{em}} \approx 1/137$, which is valid only at distances larger than the Compton wavelength, $R \gg 1/m$. Consequently, the solutions of the CAS equation are physically reliable only on the scale $r \gtrsim 1/m$.

To quantify when this regularization is negligible, we examine the factor \( (2ipr)^{\mu-1/2} \) at the scale \( \varkappa \equiv pR \ll 1 \). Expanding its squared modulus yields:
\begin{equation}
    \begin{split}
       |(2i\varkappa)^{\mu-1/2}|^2 &= (2\varkappa)^{2\sqrt{1/4 - \alpha_{\text{em}}^2}-1} = e^{-2\alpha_{\text{em}}^2 \ln(2\varkappa) + O(\alpha_{\text{em}}^4)}= 1 - 2\alpha_{\text{em}}^2 \ln(2\varkappa) + O(\alpha_{\text{em}}^4 \ln^2(\varkappa)).
    \end{split}
\end{equation}
This factor can be treated as constant (\( \approx 1 \)) provided \( -2\alpha_{\text{em}}^2 \ln(2\varkappa) \ll 1 \). Using (\ref{v_rel}) for the relative velocity \( v \) and the condition \( R \sim 1/m \), we find the resulting constraint on the velocity of the \( S^+S^- \) pair:
\[
\sqrt{1 - 4m^2/s} \gg e^{-1/(2\alpha_{\text{em}}^2)} \approx 10^{-4000}.
\]
This inequality is satisfied for any physically relevant velocity.

Therefore, the weak divergence at the origin can be safely ignored. The CAS factor is obtained by evaluating the wave function in Eq.~(\ref{Furry_scalar_solution}) at \( r=0 \), discarding the divergent factor \( (2ipr)^{\mu-1/2} \) (which is set to unity as shown above), and using \( F(a,b,0)=1 \). The final, regularized result is:
\begin{equation}
\begin{split}
    \mathcal{K}^{(\text{CAS})} &= \Bigg|\frac{\Gamma(a)}{\Gamma(b)} \cdot e^{\pi\alpha_{\text{em}}/(2v)}\Bigg|^2 = \Bigg|\frac{\Gamma(a)}{\Gamma(b)}\Bigg|^2 \cdot e^{\pi\alpha_{\text{em}}/v} = \left| \frac{ \Gamma\left( \sqrt{\frac{1}{4} - \alpha_{\text{em}}^2} + \frac{1}{2} + i\frac{\alpha_{\text{em}}}{v} \right) }{ \Gamma\left( \sqrt{1 - 4\alpha_{\text{em}}^2} + 1 \right) } \right|^2 \cdot e^{\pi\alpha_{\text{em}}/v}.
\end{split}
\end{equation}

\section{Full soft-photon QED corrections}
\label{full_QED_corr}

This appendix presents the results of the full soft-photon QED correction given by Eq.~(\ref{Isidori_factor}). For the purely leptonic decays $B_{d,s}^0 \to \ell^+\ell^-$, Eq.~(\ref{Isidori_factor}) is applied directly. For all other channels (semileptonic and radiative leptonic), the results are obtained via the average:

\begin{equation}
    \label{mean_KC_corr}
    \langle\mathcal{K}_\text{C}\rangle = \frac{\displaystyle\int\frac{d\mathcal{B}^{(\text{Coulomb})}}{d\hat{s}}\, d\hat{s}}{\displaystyle\int\frac{d\mathcal{B}^{(\text{free})}}{d\hat{s}}\, d\hat{s}} =  \frac{\displaystyle\int\frac{d\mathcal{B}^{(\text{free})}}{d\hat{s}}\cdot\mathcal{K}_\text{C}(\hat{s})\, d\hat{s}}{\displaystyle\int\frac{d\mathcal{B}^{(\text{free})}}{d\hat{s}}\, d\hat{s}}.
\end{equation}
\begin{equation}
    \label{mean_Kfull_corr}
    \langle\mathcal{K}_\text{QED}\rangle = \frac{\displaystyle\int\frac{d\mathcal{B}^{(\text{Coulomb+soft})}}{d\hat{s}}\, d\hat{s}}{\displaystyle\int\frac{d\mathcal{B}^{(\text{free})}}{d\hat{s}}\, d\hat{s}} =  \frac{\displaystyle\int\frac{d\mathcal{B}^{(\text{free})}}{d\hat{s}}\cdot\mathcal{K}_\text{QED}(\hat{s})\, d\hat{s}}{\displaystyle\int\frac{d\mathcal{B}^{(\text{free})}}{d\hat{s}}\, d\hat{s}}.
\end{equation}

The numerical results for the full soft-photon QED factors are organised as follows:
\begin{itemize}
    \item Table~\ref{tab:Bll_full_QED} lists the Coulomb factors $\mathcal{K}_C$ and the full QED factors $\mathcal{K}_{\text{QED}}$ for the purely leptonic decays $B_{d,s}^0\to\ell^+\ell^-$ ($\ell=e,\mu,\tau$) at three values of the photon energy cutoff $\Delta E$.
    \item Table~\ref{tab:Bhll_full_QED} presents the phase-space averaged factors $\langle\mathcal{K}_{\text{QED}}\rangle$ and $\langle\mathcal{K}_{\text{C}}\rangle$ for semileptonic decays with pseudoscalar mesons: $B^0\rightarrow \{K^0, \pi^0\} \ell^+\ell^-$ and $B_s^0\rightarrow \{\eta, \eta', K^0\} \ell^+\ell^-$.
    \item Table~\ref{tab:BVll_full_QED} gives the same quantities for semileptonic decays with vector mesons: $B^0\rightarrow \{K^{0*}, \rho^{0}\} \ell^+\ell^-$ and $B_s^0\rightarrow \{\phi, K^{0*}\} \ell^+\ell^-$.
    \item Table~\ref{tab:Bgammall_full_QED} contains the results for radiative leptonic decays $B_{d,s}^0\to\gamma\ell^+\ell^-$.
\end{itemize}

Several general observations can be made. First, $\mathcal{K}_{\text{QED}}$ increases with $\Delta E$: a larger cutoff allows more real radiation to be treated as undetected, which suppresses the branching fraction less strongly, so $\mathcal{K}_{\text{soft}}$ approaches unity from below. As a result, $\mathcal{K}_{\text{QED}}$ also rises with $\Delta E$.

Second, for decays into electrons or muons, the full QED correction is typically below unity: the soft‑photon suppression dominates over the Coulomb enhancement. For $\tau^+\tau^-$ final states, the Coulomb factor is larger and the soft suppression is weaker, so $\mathcal{K}_{\text{QED}}>1$.

As discussed in the Introduction, the soft part $\mathcal{K}_{\text{soft}}$ is simulated by \textsc{photos}, while the Coulomb part $\mathcal{K}_{\text{C}}$ is not. Experimental branching fractions are usually presented after applying \textsc{photos} to unfold the soft radiation (i.e. they correspond to the non‑radiative width with $\mathcal{K}_{\text{soft}}=1$). Therefore, when comparing theory with experiment, only the Coulomb correction $\mathcal{K}_{\text{C}}$ (as in Tables~\ref{tab:branching_ratio_Bll}–\ref{tab:branching_ratio_Bgammall}) should be included.

\begin{table}[ht]
\centering
\begin{tabular}{ || l | c | c | c | c || }
\hline
Decay & $\mathcal{K}_C$ & $\mathcal{K}_{\text{QED}}(0.05\ \text{GeV})$ & $\mathcal{K}_{\text{QED}}(0.1\ \text{GeV})$ & $\mathcal{K}_{\text{QED}}(0.5\ \text{GeV})$ \\
 & & & & \\ \hline
$B_s^0 \to \mu^+\mu^-$ & $1.0231$ & $0.9514$ & $0.9635$ & $0.9922$ \\
 & & & & \\
$B^0 \to \mu^+\mu^-$   & $1.0231$ & $0.9520$ & $0.9640$ & $0.9926$ \\
 & & & & \\ \hline
 $B_s^0 \to e^+e^-$ & $1.0231$ & $0.8648$ & $0.8905$ & $0.9531$ \\
 & & & & \\
$B^0 \to e^+e^-$   & $1.0231$ & $0.8611$ & $0.8874$ & $0.9517$ \\
 & & & & \\ \hline
$B_s^0 \to \tau^+\tau^-$ & $1.0241$ & $1.0051$ & $1.0084$ & $1.0160$ \\
 & & & & \\
$B^0 \to \tau^+\tau^-$   & $1.0242$ & $1.0056$ & $1.0088$ & $1.0163$ \\
\hline
\end{tabular}
\caption{Full QED factors $\mathcal{K}_{\text{QED}} = \mathcal{K}_{\text{soft}} \cdot \mathcal{K}_C$ for the decays $B_{d,s}^0 \to \ell^+\ell^-$ at different soft photon energy cutoffs $\Delta E = \{0.05, 0.1, 0.5\}$ GeV.}
\label{tab:Bll_full_QED}
\end{table}

\begin{table}[ht]
\centering
\begin{tabular}{ || l | c | c | c | c || }
\hline
Decay & $\langle\mathcal{K}_{\text{C}}\rangle$ & $\langle\mathcal{K}_{\text{QED}}\rangle(0.05\ \text{GeV})$ & $\langle\mathcal{K}_{\text{QED}}\rangle(0.1\ \text{GeV})$ & $\langle\mathcal{K}_{\text{QED}}\rangle(0.5\ \text{GeV})$ \\
\hline
$B^0 \to K^0 e^+e^-$ & $1.0233$ & $0.8754$ & $0.8996$ & $0.9584$ \\
 & & & & \\
$B^0 \to K^0 \mu^+\mu^-$ & $1.0233$ & $0.9657$ & $0.9755$ & $0.9987$ \\
 & & & & \\
$B^0 \to K^0 \tau^+\tau^-$ & $1.0334$ & $1.0206$ & $1.0228$ & $1.0280$ \\
 & & & & \\ \hline
$B^0 \to \pi^0 e^+e^-$ & $1.0233$ & $0.8735$ & $0.8980$ & $0.9575$ \\
 & & & & \\
$B^0 \to \pi^0 \mu^+\mu^-$ & $1.0233$ & $0.9636$ & $0.9738$ & $0.9978$ \\
 & & & & \\
$B^0 \to \pi^0 \tau^+\tau^-$ & $1.0303$ & $1.0164$ & $1.0188$ & $1.0245$ \\
 & & & & \\ \hline
$B_s^0 \to \eta e^+e^-$ & $1.0233$ & $0.8743$ & $0.8986$ & $0.9575$ \\
 & & & & \\
$B_s^0 \to \eta \mu^+\mu^-$ & $1.0233$ & $0.9649$ & $0.9748$ & $0.9982$ \\
 & & & & \\
$B_s^0 \to \eta \tau^+\tau^-$ & $1.0331$ & $1.0201$ & $1.0223$ & $1.0276$ \\
 & & & & \\ \hline
$B_s^0 \to \eta' e^+e^-$ & $1.0233$ & $0.8761$ & $0.9000$ & $0.9583$ \\
 & & & & \\
$B_s^0 \to \eta' \mu^+\mu^-$ & $1.0233$ & $0.9667$ & $0.9763$ & $0.9990$ \\
 & & & & \\
$B_s^0 \to \eta' \tau^+\tau^-$ & $1.0385$ & $1.0266$ & $1.0286$ & $1.0334$ \\
 & & & & \\ \hline
$B_s^0 \to K^0 e^+e^-$ & $1.0233$ & $0.8739$ & $0.8982$ & $0.9574$ \\
 & & & & \\
$B_s^0 \to K^0 \mu^+\mu^-$ & $1.0233$ & $0.9645$ & $0.9745$ & $0.9980$ \\
 & & & & \\
$B_s^0 \to K^0 \tau^+\tau^-$ & $1.0326$ & $1.0194$ & $1.0217$ & $1.0270$ \\
\hline
\end{tabular}
\caption{Full QED $\langle\mathcal{K}_{\text{QED}}\rangle$ and Coulomb  $\langle\mathcal{K}_{\text{C}}\rangle$ factors for semileptonic decays $B_{d,s}^0 \to h^0 \ell^+\ell^-$ with a pseudoscalar meson $h^0$ at different soft photon energy cutoffs $\Delta E = \{0.05, 0.1, 0.5\}$ GeV.}
\label{tab:Bhll_full_QED}
\end{table}

\begin{table}[ht]
\centering
\begin{tabular}{ || l | c | c | c | c || }
\hline
Decay & $\langle\mathcal{K}_{\text{C}}\rangle$ & $\langle\mathcal{K}_{\text{QED}}\rangle(0.05\ \text{GeV})$ & $\langle\mathcal{K}_{\text{QED}}\rangle(0.1\ \text{GeV})$ & $\langle\mathcal{K}_{\text{QED}}\rangle(0.5\ \text{GeV})$ \\
\hline
$B^0 \to K^{0*} e^+e^-$ & $1.0233$ & $0.8989$ & $0.9194$ & $0.9690$ \\
 & & & & \\
$B^0 \to K^{0*} \mu^+\mu^-$ & $1.0234$ & $0.9676$ & $0.9771$ & $0.9996$ \\
 & & & & \\
$B^0 \to K^{0*} \tau^+\tau^-$ & $1.0366$ & $1.0244$ & $1.0265$ & $1.0315$ \\
 & & & & \\ \hline
$B^0 \to \rho^0 e^+e^-$ & $1.0234$ & $0.9134$ & $0.9315$ & $0.9754$ \\
 & & & & \\
$B^0 \to \rho^0 \mu^+\mu^-$ & $1.0235$ & $0.9696$ & $0.9788$ & $1.0005$ \\
 & & & & \\
$B^0 \to \rho^0 \tau^+\tau^-$ & $1.0338$ & $1.0212$ & $1.0234$ & $1.0285$ \\
 & & & & \\ \hline
$B_s^0 \to \phi e^+e^-$ & $1.0234$ & $0.9207$ & $0.9377$ & $0.9785$ \\
 & & & & \\
$B_s^0 \to \phi \mu^+\mu^-$ & $1.0236$ & $0.9728$ & $0.9815$ & $1.0018$ \\
 & & & & \\
$B_s^0 \to \phi \tau^+\tau^-$ & $1.0370$ & $1.0249$ & $1.0270$ & $1.0319$ \\
 & & & & \\ \hline
$B_s^0 \to K^{0*} e^+e^-$ & $1.0234$ & $0.9167$ & $0.9342$ & $0.9767$ \\
 & & & & \\
$B_s^0 \to K^{0*} \mu^+\mu^-$ & $1.0235$ & $0.9712$ & $0.9801$ & $1.0011$ \\
 & & & & \\
$B_s^0 \to K^{0*} \tau^+\tau^-$ & $1.0352$ & $1.0227$ & $1.0248$ & $1.0299$ \\ \hline
\end{tabular}
\caption{Full QED $\langle\mathcal{K}_{\text{QED}}\rangle$ and Coulomb  $\langle\mathcal{K}_{\text{C}}\rangle$ factors for semileptonic decays $B_{d,s}^0 \to V^0 \ell^+\ell^-$ with a vector meson $V^0$ at different soft photon energy cutoffs $\Delta E = \{0.05, 0.1, 0.5\}$ GeV.}
\label{tab:BVll_full_QED}
\end{table}

\begin{table}[ht]
\centering
\begin{tabular}{ || l | c | c | c | c || }
\hline
$q^2\in[1,6]\ \text{GeV}^2$ & $\langle\mathcal{K}_{\text{C}}\rangle$ & $\langle\mathcal{K}_{\text{QED}}\rangle(0.05\ \text{GeV})$ & $\langle\mathcal{K}_{\text{QED}}\rangle(0.1\ \text{GeV})$ & $\langle\mathcal{K}_{\text{QED}}\rangle(0.5\ \text{GeV})$ \\
\hline
$B^0 \to \gamma e^+e^-$ & $1.023$ & $0.881$ & $0.905$ & $0.961$ \\
 & & & & \\ 
$B^0 \to \gamma \mu^+\mu^-$ & $1.023$ & $0.972$ & $0.981$ & $1.002$ \\
 & & & & \\ \hline
$B_s^0 \to \gamma e^+e^-$ & $1.023$ & $0.889$ & $0.911$ & $0.964$ \\
 & & & & \\ 
$B_s^0 \to \gamma \mu^+\mu^-$ & $1.023$ & $0.981$ & $0.988$ & $1.005$ \\
 & & & & \\ \hline \hline
$q^2\in[4m_\tau^2, M_{B_{d,s}}^2]$ & $\langle\mathcal{K}_{\text{C}}\rangle$ & $\langle\mathcal{K}_{\text{QED}}\rangle(0.05\ \text{GeV})$ & $\langle\mathcal{K}_{\text{QED}}\rangle(0.1\ \text{GeV})$ & $\langle\mathcal{K}_{\text{QED}}\rangle(0.5\ \text{GeV})$ \\ \hline
$B^0 \to \gamma \tau^+\tau^-$ & $1.027$ & $1.012$ & $1.015$ & $1.021$ \\
 & & & & \\ 
$B_s^0 \to \gamma \tau^+\tau^-$ & $1.027$ & $1.011$ & $1.014$ & $1.020$ \\ \hline
\end{tabular}
\caption{Full QED $\langle\mathcal{K}_{\text{QED}}\rangle$ and Coulomb  $\langle\mathcal{K}_{\text{C}}\rangle$ factors for radiative leptonic decays $B_{d,s}^0 \to \gamma \ell^+\ell^-$ at different soft photon energy cutoffs $\Delta E = \{0.05, 0.1, 0.5\}$ GeV. The integration ranges are $q^2\in[1,6]$ GeV$^2$ for $\ell = e,\mu$ and $q^2\in[4m_\tau^2, M_{B_{d,s}}^2]$ for $\ell = \tau$.}
\label{tab:Bgammall_full_QED}
\end{table}

\FloatBarrier
\section{Differential, angle and double differential distributions}
\label{app_plots}

\begin{figure*}[h]
\centering
\includegraphics[width=\linewidth]{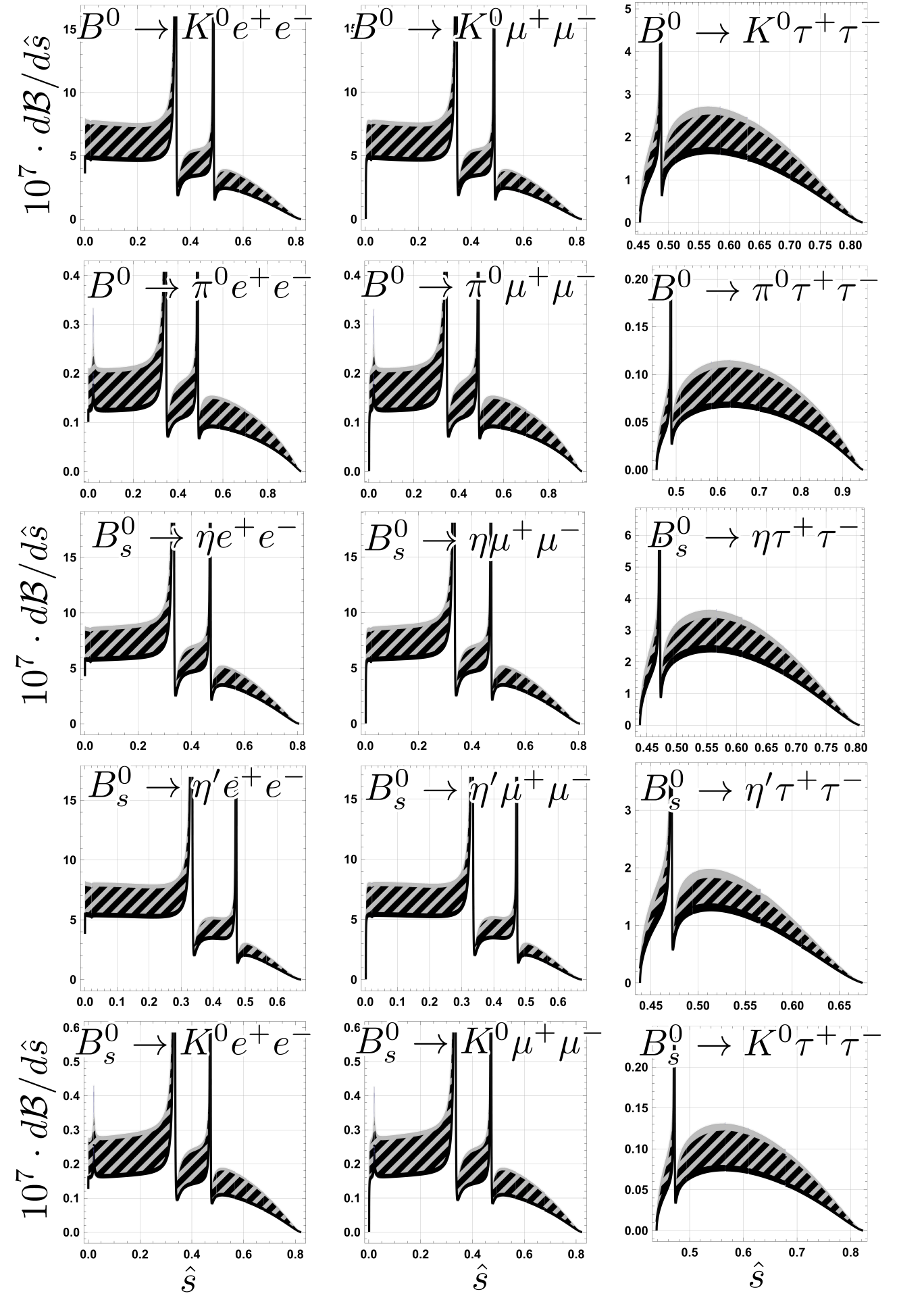}
\caption{Dependence of the differential branching fraction $10^7 d\mathcal{B}/d\hat{s}$ for the decays $B^0\rightarrow \{K^0, \pi^0\} \ell^+\ell^-$ and $B_s^0\rightarrow \{\eta, \eta', K^0\} \ell^+\ell^-$ on $\hat{s} = (p_{B_{d,s}^0}-p_{h^0})^2/M_{B_{d,s}^0}^2$ — the squared transferred momentum normalized to the square of the $B$-meson mass. The black band corresponds to predictions without Coulomb interaction, the gray band — with Coulomb interaction. The overlap region is indicated by black-gray hatching.}
\label{pic_Bhll}
\end{figure*}

\begin{figure*}[h]
\centering
\includegraphics[width=\linewidth]{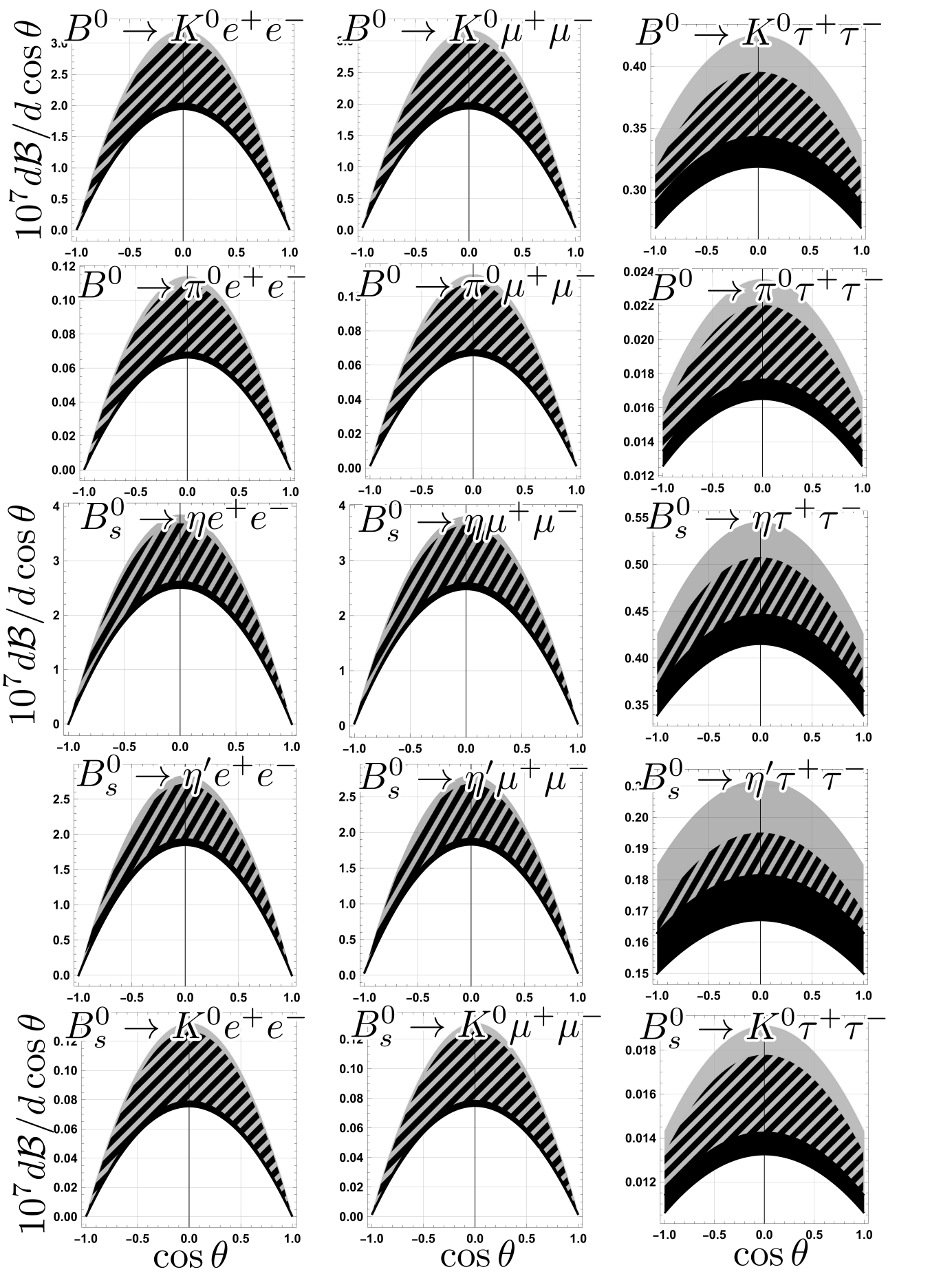}
\caption{Angular distributions $10^7 d\mathcal{B}/d\cos\theta$ for the decays $B^0\rightarrow \{K^0, \pi^0\} \ell^+\ell^-$ and $B_s^0\rightarrow \{\eta, \eta', K^0\} \ell^+\ell^-$ as a function of $\cos\theta$, where $\theta = \angle (\mathbf{p} _{h^0}, \mathbf{p}_{\ell^+})$ is the angle between the direction of the neutral hadron $h^0$ and the positive lepton $\ell^+$ in the $\ell^+\ell^-$ rest frame. The black band corresponds to predictions without Coulomb interaction, the gray band — with Coulomb interaction. The overlap region is indicated by black-gray hatching.}
\label{pic_Bhll_angle}
\end{figure*}

\begin{figure*}[h]
\centering
\includegraphics[width=\linewidth]{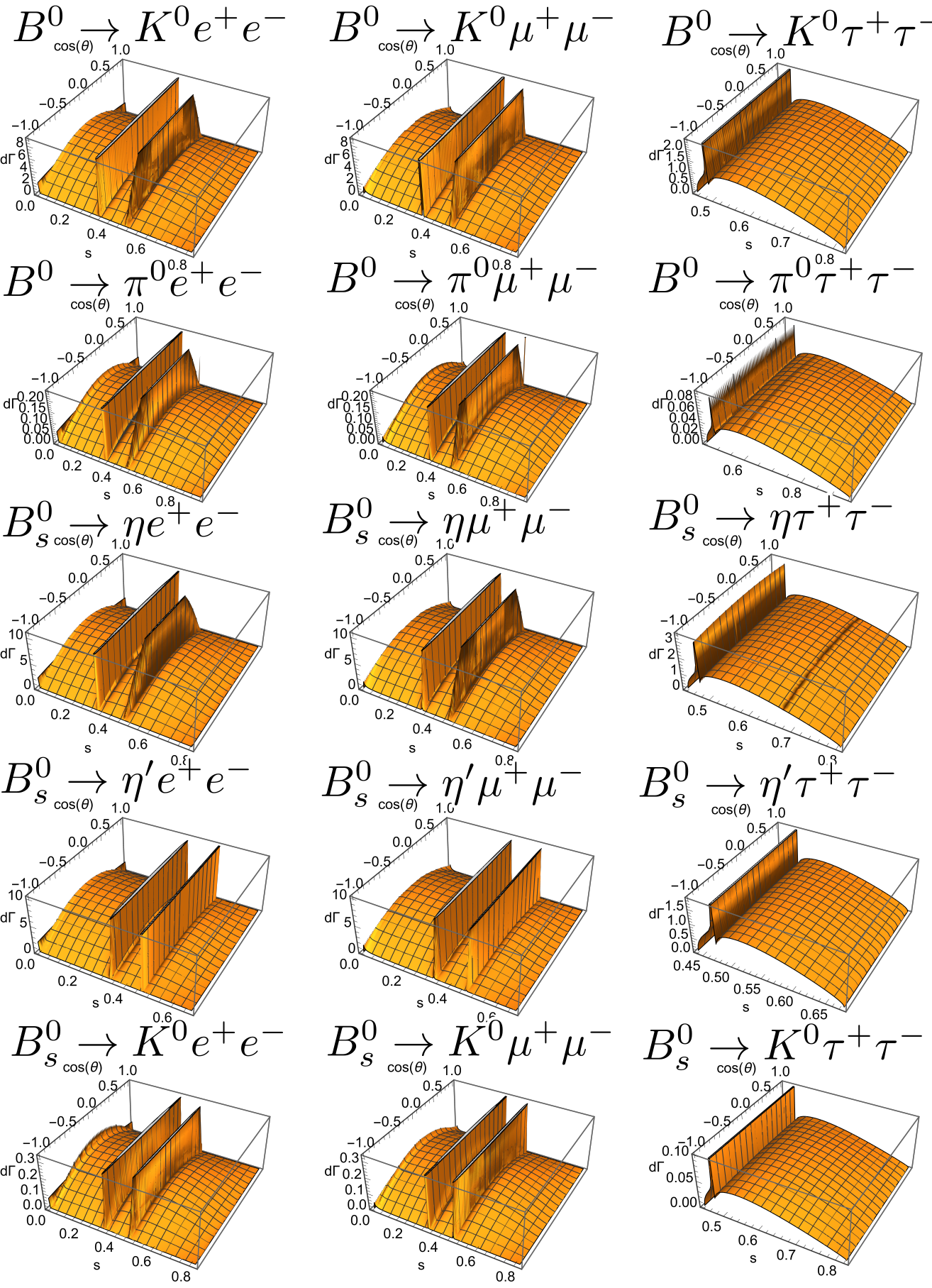}
\caption{Double differetial distributions $10^7\cdot d\mathcal{B}/d\hat{s}d\cos\theta$ for the decays  $B^0\rightarrow \{K^0, \pi^0\} \ell^+\ell^-$ and $B_s^0\rightarrow \{\eta, \eta', K^0\} \ell^+\ell^-$. Here $\hat{s} = (p_{B_{d,s}^0}-p_{h^0})^2/M_{B_{d,s}^0}^2$ and $\theta = \angle (\mathbf{p} _{h^0}, \mathbf{p}_{\ell^+})$ is the angle between the direction of the neutral hadron $h^0$ and the positive lepton $\ell^+$ in the $\ell^+\ell^-$ rest frame. }
\label{pic_Bhll_3D}
\end{figure*}

\begin{figure*}[h]
\centering
\includegraphics[width=\linewidth]{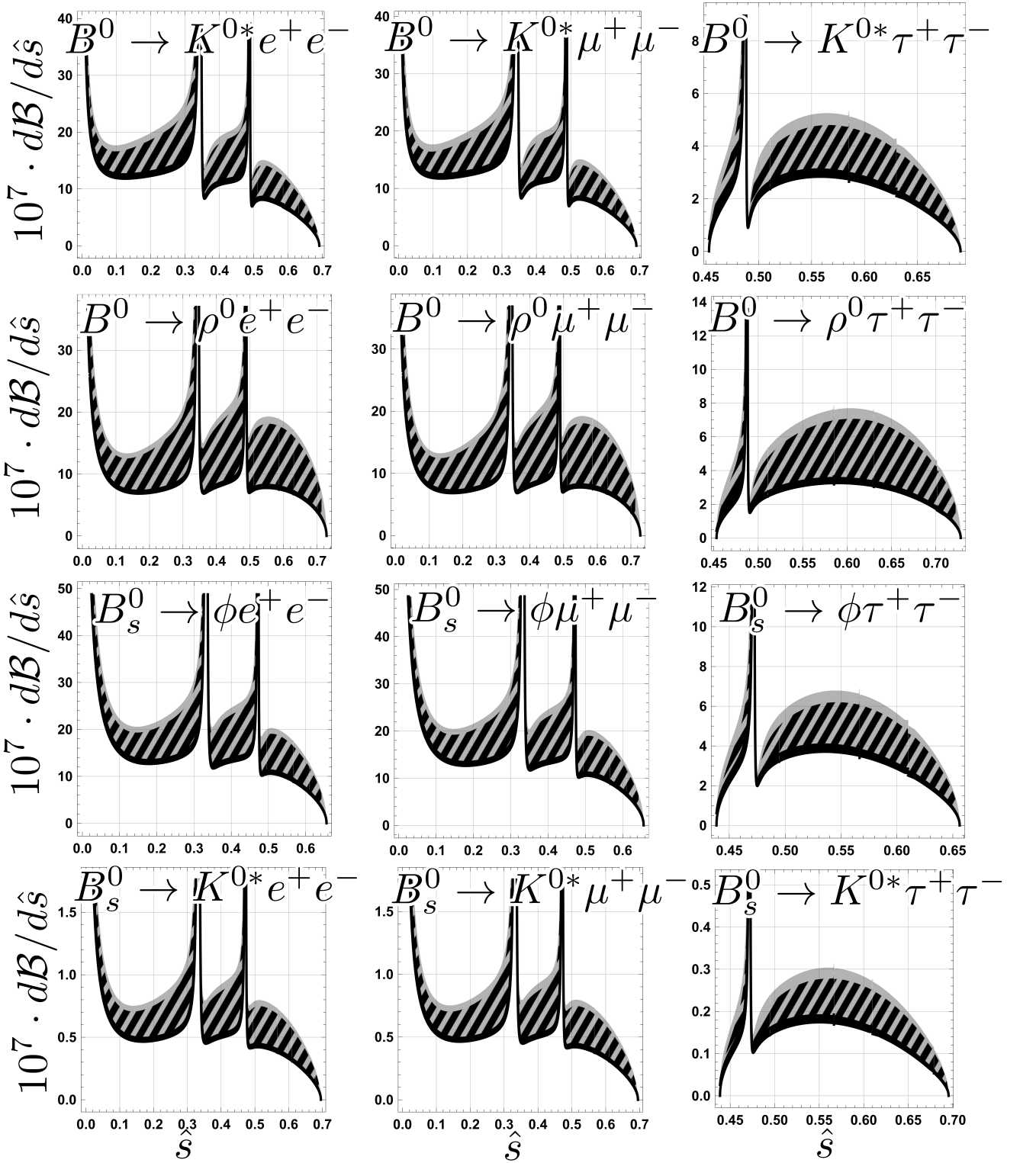}
\caption{Dependence of the differential branching fraction $10^7 d\mathcal{B}/d\hat{s}$ for the decays $B^0\rightarrow \{K^{0*}, \rho^0\} \ell^+\ell^-$ and $B_s^0\rightarrow \{\phi, K^{0*}\} \ell^+\ell^-$ on $\hat{s} = (p_{B_{d,s}^0}-p_{V^0})^2/M_{B_{d,s}^0}^2$ — the squared transferred momentum normalized to the square of the $B$-meson mass. The black band corresponds to predictions without Coulomb interaction, the gray band — with Coulomb interaction. The overlap region is indicated by black-gray hatching.}
\label{pic_BVll}
\end{figure*}

\begin{figure*}[h]
\centering
\includegraphics[width=\linewidth]{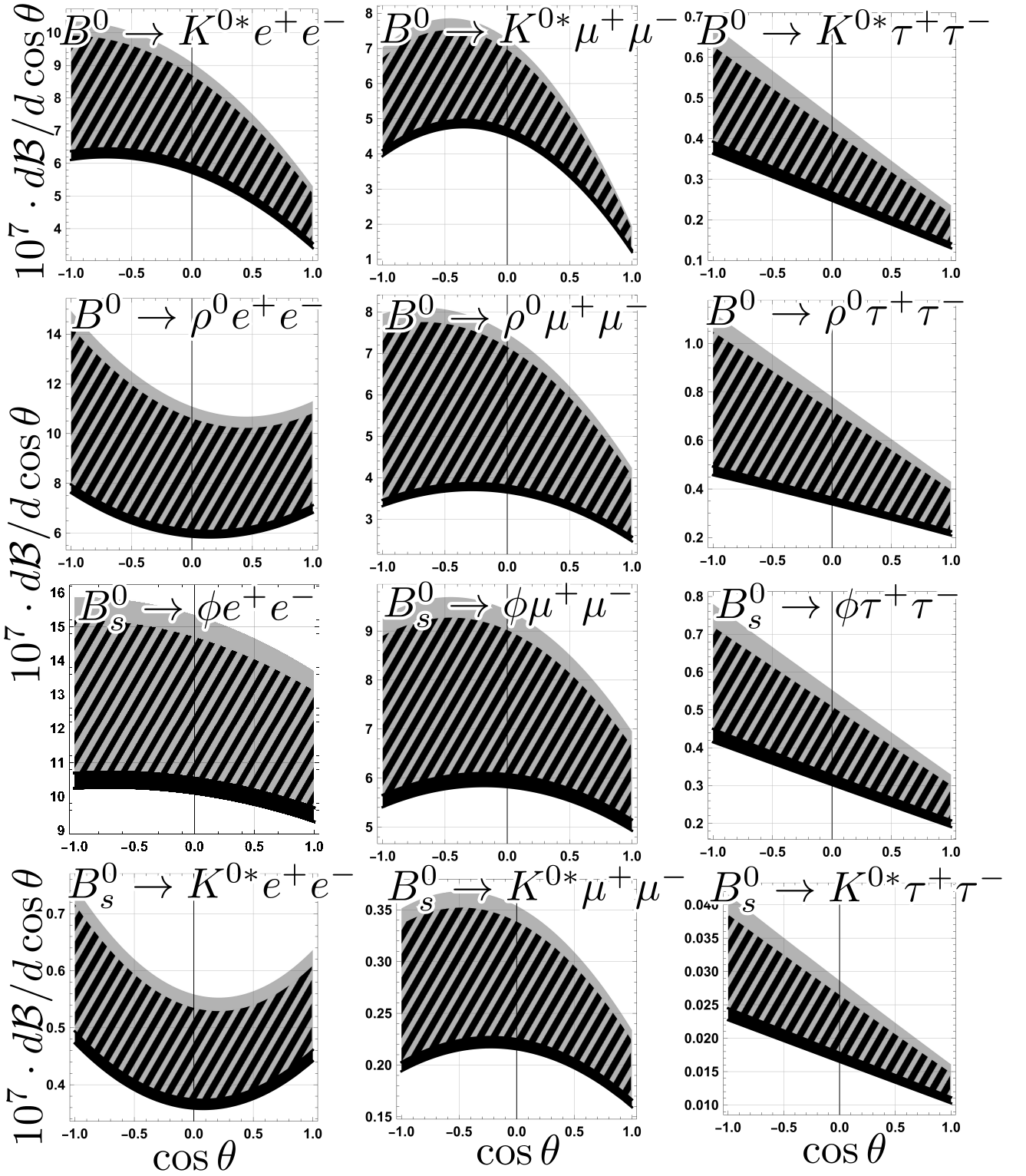}
\caption{Angular distributions for $B^0\rightarrow \{K^{0*}, \rho^0\} \ell^+\ell^-$ and $B_s^0\rightarrow \{\phi, K^{0*}\} \ell^+\ell^-$ as a function of $\cos\theta$, where $\theta = \angle (\mathbf{p} _{V^0}, \mathbf{p}_{\ell^+})$ is the angle between the direction of the neutral hadron $V^0$ and the positive lepton $\ell^+$ in the $\ell^+\ell^-$ rest frame. The black band corresponds to predictions without Coulomb interaction, the gray band — with Coulomb interaction. The overlap region is indicated by black-gray hatching.}
\label{pic_BVll_angle}
\end{figure*}

\begin{figure*}[h]
\centering
\includegraphics[width=\linewidth]{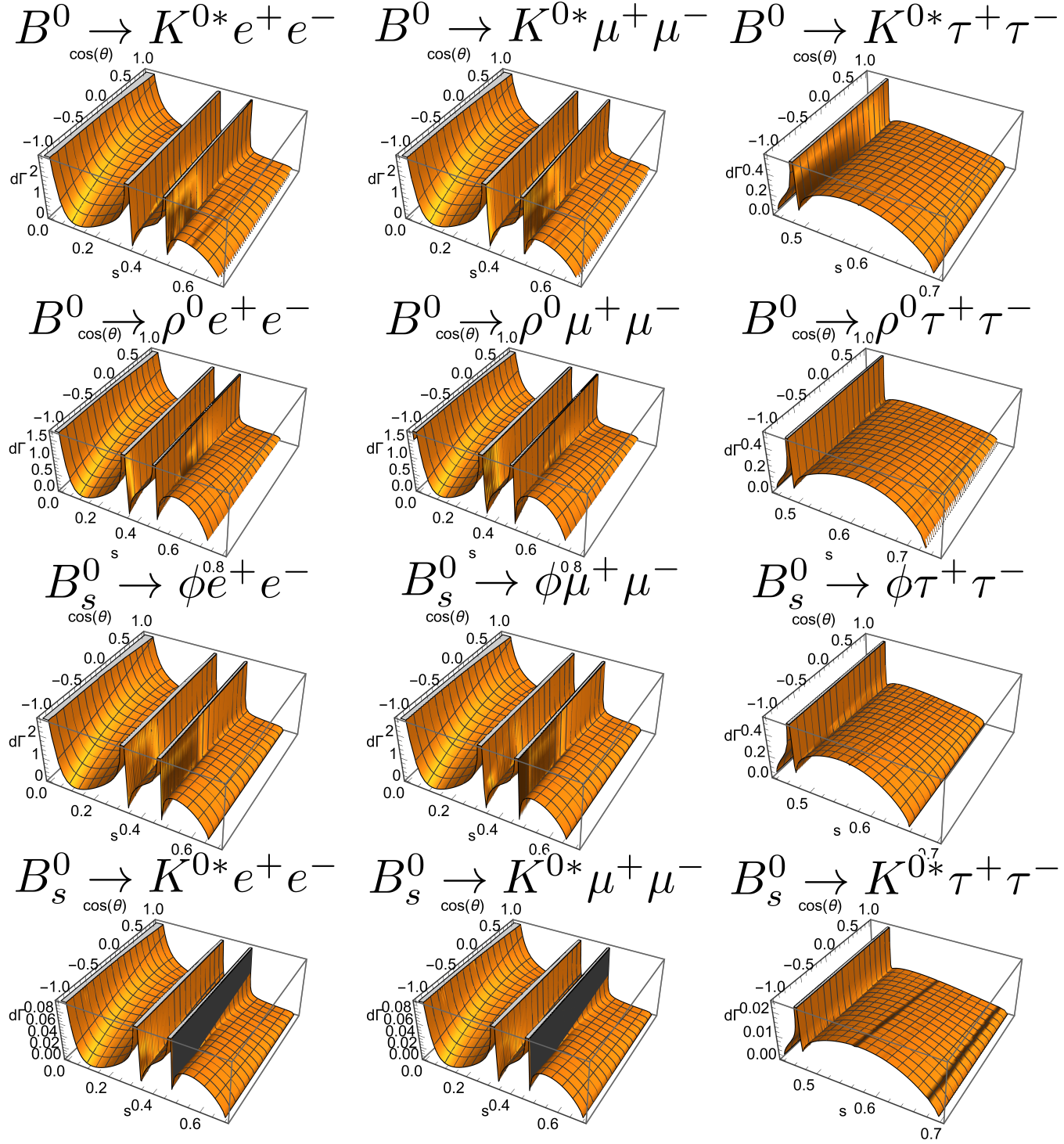}
\caption{Double differetial distributions $10^7\cdot d\mathcal{B}/d\hat{s}d\cos\theta$ for $B^0\rightarrow \{K^{0*}, \rho^0\} \ell^+\ell^-$ and $B_s^0\rightarrow \{\phi, K^{0*}\} \ell^+\ell^-$. Here $\hat{s} = (p_{B_{d,s}^0}-p_{V^0})^2/M_{B_{d,s}^0}^2$ and $\theta = \angle (\mathbf{p} _{V^0}, \mathbf{p}_{\ell^+})$ is the angle between the direction of the neutral hadron $V^0$ and the positive lepton $\ell^+$ in the $\ell^+\ell^-$ rest frame.}
\label{pic_BVll_3D}
\end{figure*}

\begin{figure*}[h]
\centering
\includegraphics[width=\linewidth]{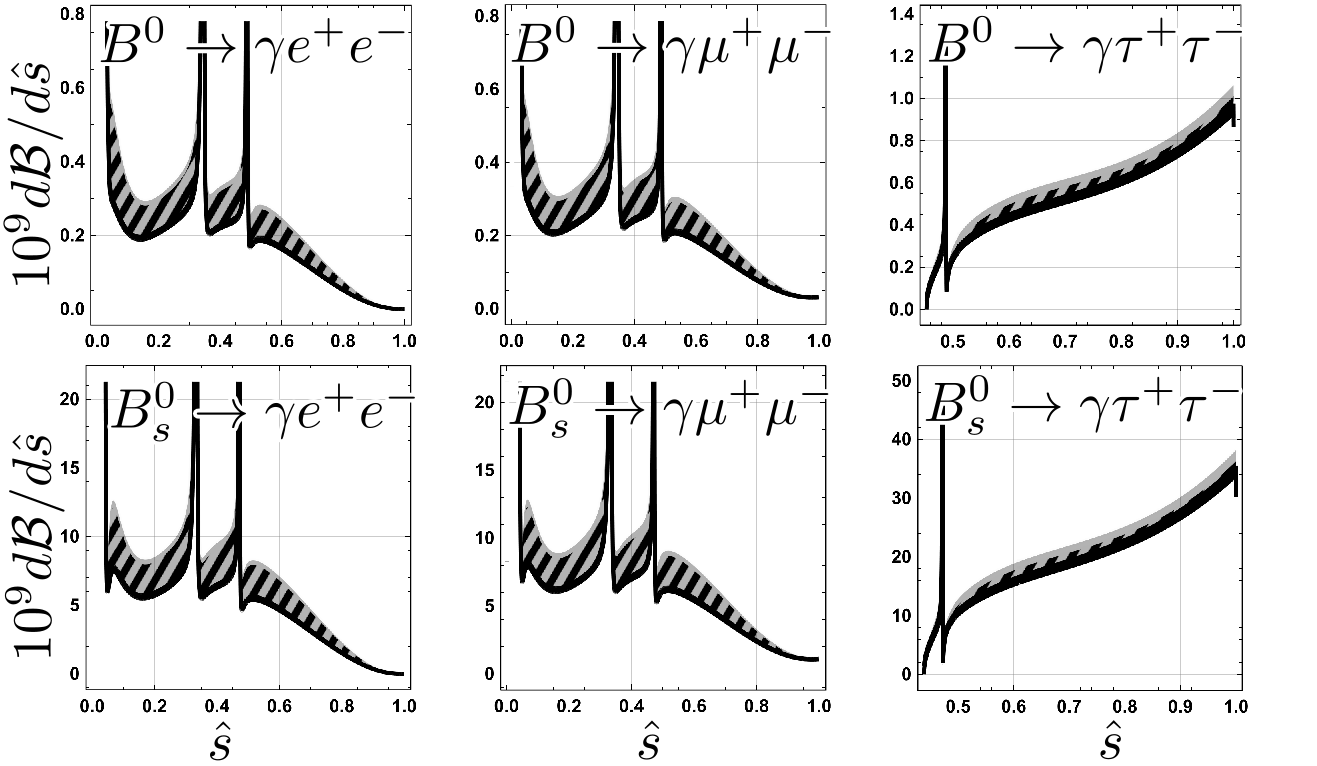}
\caption{Dependence of the differential branching fraction $10^9 d\mathcal{B}/d\hat{s}$ for the decays $B_{d,s}^0\rightarrow \gamma \ell^+\ell^-$ on $\hat{s} = (p_{B_{d,s}^0}-p_\gamma)^2/M_{B_{d,s}^0}^2$. The black band corresponds to predictions without Coulomb interaction, the gray band — with Coulomb interaction. The overlap region is indicated by black-gray hatching.}
\label{pic_Bgammall}
\end{figure*}

\begin{figure*}[h]
\centering
\includegraphics[width=\linewidth]{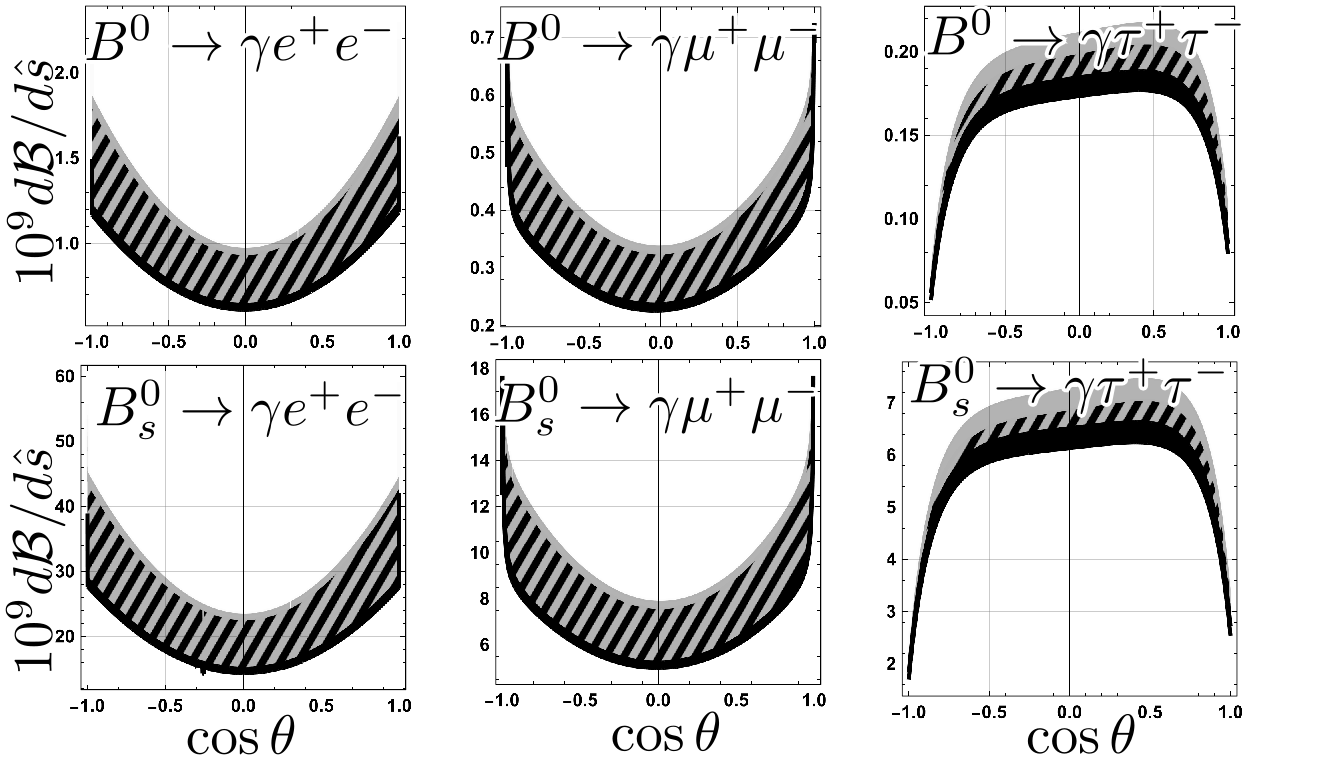}
\caption{Angular distributions for $B_{d,s}^0\rightarrow \gamma \ell^+\ell^-$ as a function of $\cos\theta$, where $\theta = \angle (\mathbf{p} _{\gamma}, \mathbf{p}_{\ell^+})$ is the angle between the direction of the photon $\gamma$ and the positive lepton $\ell^+$ in the $\ell^+\ell^-$ rest frame. The black band corresponds to predictions without Coulomb interaction, the gray band — with Coulomb interaction. The overlap region is indicated by black-gray hatching.}
\label{pic_Bgammall_angle}
\end{figure*}

\begin{figure*}[h]
\centering
\includegraphics[width=\linewidth]{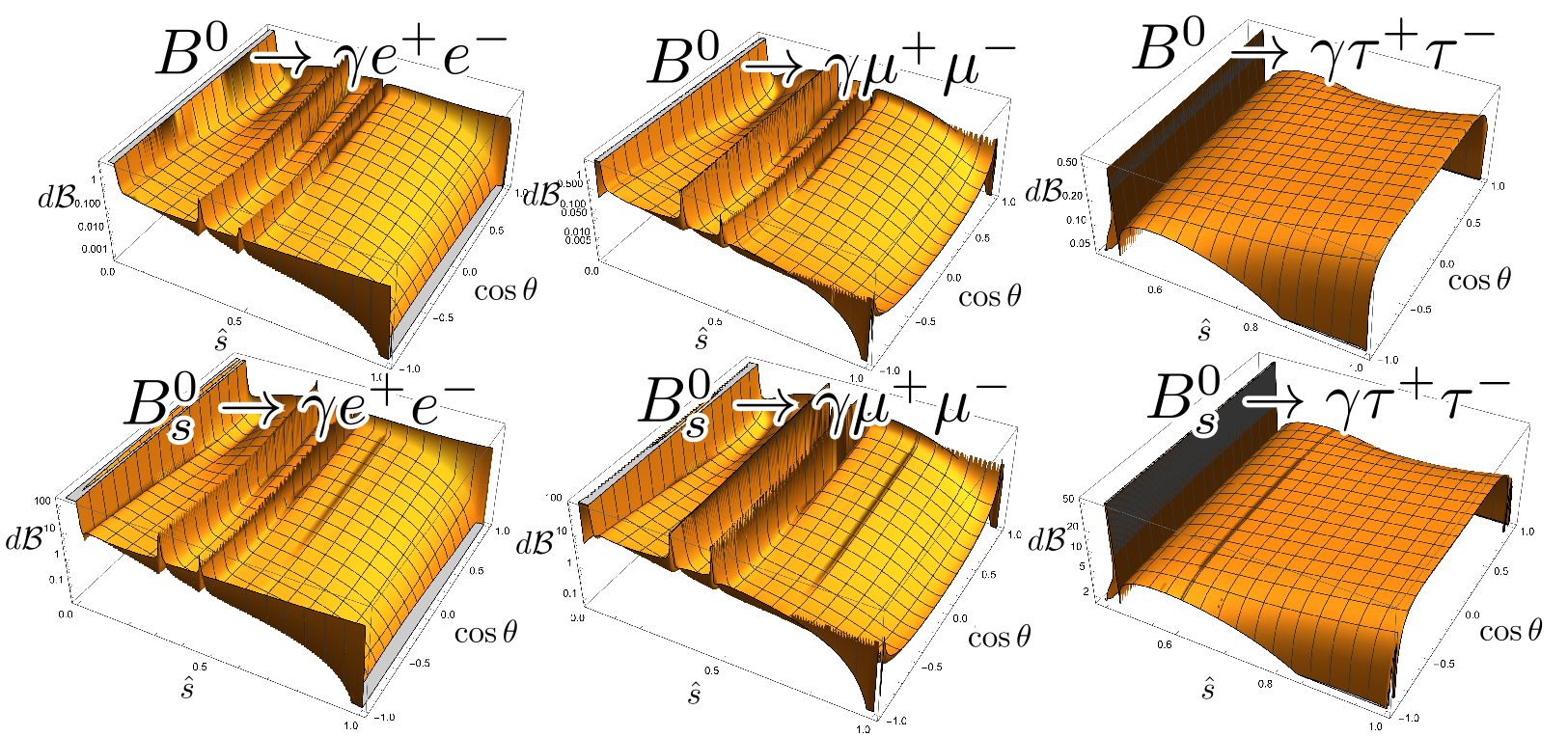}
\caption{Double differetial distributions $10^9\cdot d\mathcal{B}/d\hat{s}d\cos\theta$ for $B_{d,s}^0\rightarrow \gamma \ell^+\ell^-$. Here $\hat{s} = (p_{B_{d,s}^0}-p_\gamma)^2/M_{B_{d,s}^0}^2$ and $\theta = \angle (\mathbf{p} _{\gamma}, \mathbf{p}_{\ell^+})$ is the angle between the direction of the photon $\gamma$ and the positive lepton $\ell^+$ in the $\ell^+\ell^-$ rest frame.}
\label{pic_Bgammall_3D}
\end{figure*}

This appendix contains the complete set of differential, angular, and double differential distributions for all decay channels studied in this work. The figures are organized as follows:

\begin{itemize}
    \item Fig.~\ref{pic_Bhll}: Differential distributions $d\mathcal{B}/d\hat{s}$ for decays with pseudoscalar mesons $B^0_{d,s}\to h^0\ell^+\ell^-$.
    \item Fig.~\ref{pic_Bhll_angle}: Angular distributions $d\mathcal{B}/d\cos\theta$ for the same pseudoscalar channels.
    \item Fig.~\ref{pic_Bhll_3D}: Double differential distributions $d\mathcal{B}/d\hat{s}d\cos\theta$ for pseudoscalar meson decays.
    \item Fig.~\ref{pic_BVll}: Differential distributions $d\mathcal{B}/d\hat{s}$ for decays with vector mesons $B^0_{d,s}\to V^0\ell^+\ell^-$.
    \item Fig.~\ref{pic_BVll_angle}: Angular distributions $d\mathcal{B}/d\cos\theta$ for vector meson channels.
    \item Fig.~\ref{pic_BVll_3D}: Double differential distributions $d\mathcal{B}/d\hat{s}d\cos\theta$ for vector meson decays.
    \item Fig.~\ref{pic_Bgammall}: Differential distributions $d\mathcal{B}/d\hat{s}$ for radiative leptonic decays $B^0_{d,s}\to \gamma\ell^+\ell^-$.
    \item Fig.~\ref{pic_Bgammall_angle}: Angular distributions $d\mathcal{B}/d\cos\theta$ for radiative leptonic decays.
    \item Fig.~\ref{pic_Bgammall_3D}: Double differential distributions $d\mathcal{B}/d\hat{s}d\cos\theta$ for radiative leptonic decays.
\end{itemize}

Here $\hat{s} = (p_{B_{d,s}^0} - p_X)^2 / M_{B_{d,s}^0}^2$ denotes the squared transferred four-momentum normalized to the square of the ${B_{d,s}^0}$-meson mass, where $X$ stands for the corresponding final-state particle ($h^0$, $V^0$ or $\gamma$) and
the angular variable $\cos\theta$ is defined via the angle $\theta$ between the momentum of the particle $X$ and the momentum of the positive lepton $\ell^+$ in the dilepton ($\ell^+\ell^-$) rest frame.

In all plots, the black band represents theoretical predictions without Coulomb interaction, the gray band includes Coulomb corrections, and the overlapping region is shown with black-gray hatching.

\end{document}